\RequirePackage[l2tabu,orthodox]{nag}
\documentclass[3p]{elsarticle}

\usepackage{framed}
\usepackage{amsmath,amssymb,amsthm}

\usepackage{natbib}

\usepackage{graphicx,color}

\definecolor{darkgreen}{rgb}{0.0,0.7,0.0}

\definecolor{darkred}{rgb}{0.75,0.0,0.0}

\newcommand{\Ar}{\,\mathrm{A}}
\newcommand{\Br}{\,\mathrm{B}}
\newcommand{\Cr}{\mathrm{C}}
\newcommand{\Er}{\,\mathrm{E}}
\newcommand{\Ir}{\,\mathrm{I}}
\newcommand{\Prm}{\mathrm{P}}
\newcommand{\Tr}{\mathrm{T}}
\newcommand{\Zr}{\,\mathrm{Z}}

\newcommand{\R}{\mathbf{R}}

\DeclareMathOperator{\adj}{adj}
\DeclareMathOperator{\col}{col}
\theoremstyle{plain}
\newtheorem{thm}{Theorem}[section]
\newtheorem{lem}[thm]{Lemma}

\theoremstyle{definition}
\newtheorem{defn}{Definition}[section]
\newtheorem{exmp}{Example}[section]
\newtheorem{rem}{Remark}[section]
\newproof{pf}{Proof}
\begin{document}

\begin{frontmatter}
\title{Within-host phenotypic evolution and the population-level control of chronic viral infections by treatment and prophylaxis}
\author[DG]{Dmitry Gromov\corref{corrauth}}
\cortext[corrauth]{Corresponding author}
\ead{dv.gromov@gmail.com}

\author[ERS]{Ethan O. Romero-Severson}\ead{eoromero@lanl.gov}

\address[DG]{Faculty of Applied Mathematics and Control Processes, Saint Petersburg State University,\\ St. Petersburg, Russia}
\address[ERS]{Theoretical Biology and Biophysics Group, Los Alamos National Laboratory, \\Los Alamos, New Mexico, USA}

\begin{abstract}
Chronic viral infections can persist in an infected person for decades. 
From the perspective of the virus, a single infection can span thousands of generations, leading to a highly diverse population of viruses with its own complex evolutionary history.
We propose a mathematical framework for understanding how the emergence of new viral strains and phenotype within infected persons affects the population-level control of those infections by both non-curative treatment and chemo-prophylactic measures.  
We consider the within-host emergence of new strains that lack phenotype novelty and also the evolution of variability in contagiousness, resistance to therapy, and efficacy of prophylaxis.
Our framework balances the need for verisimilitude with our desire to retain a model that can be approached analytically.
We show how to compute the population-level basic reproduction number accounting for the within-host evolutionary process where new phenotypes emerge and are lost in infected persons, which we also extend to include both treatment and prophylactic control efforts.
This allows us to make clear statements about both the global and relative efficacy of different control efforts accounting for within-host phenotypic evolution.    
Finally, we give expressions for the endemic equilibrium of these models for certain constrained versions of the within-host evolutionary model providing a potential method for estimating within-host evolutionary parameters from population-level genetic sequence data.   

\end{abstract}


\end{frontmatter}

\section{Introduction}
Pathogens that lead to persistent chronic infection in people must mitigate both the innate and adaptive immune systems.
Strategies for evading the innate immune system are complex including direct subversion of host signaling pathways \cite{MMS:09}.
Pathogens such as HIV avoid the adaptive immune system by simply evolving new phenotypes faster than the host immune system can adapt leading to a rapid co-evolutionary race.
Because HIV has a short generation time and generates a massive number of new viral particles each day \cite{Perelson:96}, this evolutionary race creates a large potential for the emergence of new viral strains and phenotypes.
This rapid evolutionary process is one of the many reasons that HIV is exceedingly difficult to treat.
Mutations that evolve in a single host are also known to be transmitted.
In 2014-2016, 6 out of 11 countries looking for the presence of pre-treatment drug resistance (i.e. presence of a drug resistant phenotype in persons unexposed to the drug) reported greater than 10\% of new infections were resistant to one or more non-nucleoside reverse-transcriptase inhibitor, which is related to both treatment failure and death \cite{WHO:19}. 
The rapid emergence of new viral phenotypes within infected persons is not just a clinical problem, it is an epidemiological problem.
Chemo-prophylactic measures that focus on protecting uninfected persons using similar drugs to those used for treating infected persons are not immune to evolutionary derailment.
In King Country, Washington, 0.5\% people living with HIV were found to have resistance mutations to the drugs used for prophylaxis \cite{WHO:19}. 
However, with the emergence of chemo- and bio-prophylactic agents (i.e. anti-HIV antibodies for prevention), we must consider the possibility that population-level administration of these agents can shift the ever evolving landscape of chronic viral infections.

This paper is motivated by a need for mathematical models that integrate within-host genetic diversity and phenotypic evolution with epidemiological dynamics and consider the effects of joint therapeutic and prophylactic  controls.
We also attempted to balance the complexity of the model to be usable as a data analysis tool with the desire to understand the mathematical and statistical properties of the model using analytical methods.   
Our model accounts for within-host evolution among multiple phenotypes characterized by variable contagiousness, resistance to prophylactic measures, and resistance to therapeutic measures. 
Our framework allows for new phenotypes to emerge in chronic infection that can be both transmitted and possibly lost in later hosts. 
We consider both the epidemiological and evolutionary effects of both therapy for infected persons and chemo-prophylaxis-type measures for uninfected persons.



There has been a number of results devoted to the analysis of different aspects of the evolutionary and epidemiological dynamics of a multi-strain pathogen. While there is a wide spectrum of different models covering different aspects of virus/immune system evolution and their interaction, most developed models are too complex to be analyzed analytically thus, their analysis restricts to carrying out and analyzing the results of numerical simulations. Our model is related to the approach to Lythgoe et al. \cite{Lythgoe:13} that considers the possibility of a person infected with virus of type $i$ can transmit a virus of type $j$ at a time-dependent rate $\beta_{ij}(t)$. While this approach presents a detailed model of the within-host viral evolution, it requires a substantial amount of data which is not readily available: virus reproduction, mutation and death rates. Furthermore, since we need to take into account the duration of the infection at the time when transmission occurs, the system dynamics is governed by integro-differential equations which are difficult to deal with. On the other hand, such a detailed approach turns out to be an overkill as the total pool of infected contain the individuals at different stages of disease and hence, the transmission rates undergo a sort of averaging over the whole set of infected. Therefore, we employ a simpler formalism in which we treat the virus evolutionary dynamics in a more coarse grained fashion. This allows us to balance our mutual goals of a sufficiently complex model that can still be aproached analytically.  

Complex multi-strain models have been proposed for influenza \cite{Kryazhimskiy:07,Minayev:08} and dengue \cite{Bianco:09} that focus on both cross-reactivity among circulating strains and coinfection \cite{Ventel:11} rather than the emergence of new strains within infected hosts. 
Much of this work of this work is often based on complex models that are intended to explain specific biological phenomena that are too complex to be understood by applied analysis methods. 
On the other hand, there has been a number of papers devoted to the analytic analysis of certain aspects of multi-strain virus dynamics. However, most of the papers either deal with rather restricted setups or study only certain aspects of the system dynamics. We mention stability analysis of within-host multi-strain virus dynamics with mutations \cite{Leenheer:08}; analysis of a multi-strain (actually two-strain) disease with environmental transmission, no mutations \cite{Breban:10}; bifurcation analysis of a number of (rather simple) multi-strain epidemiological models without mutation \cite{Kooi:13}.

Further information about different approaches to modeling the evolutionary and population-based dynamics of multi-strain pathogens as well as the description of the problems that arise in this connection can be found in \cite{Kucharski:16,Wikra:15}

It should be noted that most research effort aimed at studying the dynamics of multi-strain viruses does not take into account the possibility of within-host mutations and concentrate on modeling different immune system responses in reaction to re-infection or co-infection. In contrast to that, we are more concerned with the effect of mutations on both the within-host and population-level distribution of viral strains and on how both the emergence and loss of phenotypes within infected persons alters the population-level control of chronic viral infections.

\section{Description of the models and their structural properties}

\subsection{A baseline model of a chronic multi-strain virus infection}
In the baseline version of the model we consider the within-host evolution and transmission of distinct strains that have the same phenotype.
Most observed mutations at the nucleotide level will not alter a pathogen's phenotype, but they can be used in combination to discern unique lineages of transmission. 
In order to account for virus variability, the whole set of viruses is divided into $n$ groups (strains), $V_i$, $i=1,\ldots,n$. An individual gets infected by a virus of particular type $i$. During the acute stage the patient's viral population consists of viruses of the same type, while during the chronic stage the original virus mutates thus producing strains of all types. Therefore, we assume that each chronically infected individual's viral population contains viruses of all $n$ types in the proportion that depends on the initial type of the virus. 


%

To model the process of disease propagation we employ an SI model and assume that there are two stages of the disease: the acute and the chronic one. Furthermore, we extend the set of state variables to include the individuals enrolled into treatment. In doing so we assume that the treatment is completely efficient and the patients are fully compliant with the treatment. 

When writing the differential equations of the model we assume the inflow to be equal to the outflow hence, the total population size remains constant. Therefore, we write the model equations for the fractions of the respective cohorts in the total population. This implies, in particular, that the sum of all the states is equal to 1. We have the following set of DEs (state variables and parameters defined in Table \ref{tab1}): 

\begin{equation}\label{eq:sys1}\begin{aligned}
    \dot{I}_{Ai}={ }&\phi_{i}(X)S -\gamma I_{Ai}-\mu I_{Ai}\\
    \dot{I}_{Ci}={ }&\gamma I_{Ai}- u_{\Tr} I_{Ci} - \mu I_{Ci}\\
    \dot{T}= { }& u_{\Tr}\sum_{i=1}^n I_{Ci}-\mu T\\
    \dot{S}= { }&\mu - \sum_{i=1}^n\phi_{i}(X)S -\mu S
\end{aligned}\end{equation}
where $X=[I_{A1},\dots,I_{An},I_{C1},\dots,I_{Cn}, T,S]$ is the $(2n+2)$-dimensional state vector, $i$ ranges from $1$ to $n$, and the respective forces of infection are defined as
\begin{equation}\label{eq:phi1}\phi_{i}(X)=\beta_A I_{Ai}+\beta_{\Cr}\sum_{j=1}^n \alpha_{ij} I_{Cj}.\end{equation}
In \eqref{eq:phi1}, $\alpha_{ij}\in[0,1]$ denotes the average fraction of type $i$ viruses in the viral population of an individual initially infected by the type $j$ virus. It should thus hold that $\sum_{i=1}^n \alpha_{ij}=1$ for all $j=1,\dots,n$. Furthermore, we assume that $\alpha_{ii}\neq 0$ for all $i=1,\dots,n$. This means that the viral population of an individual infected with the type $i$ virus always contains a non-zero amount of the corresponding strain. Parameters $\beta_A$ and $\beta_{\Cr}$ are the transmissibility rates of acute and chronically infected individuals.  In this simple setting we assume that the probability that a susceptible individual contracts a disease depends only on the disease stage of the infected contact, but not on the type of virus. That is to say, a susceptible can be equally well infected by any virus. In the following, we will assume that $\beta_A$ and $\beta_{\Cr}$ differ by a proportionality coefficient $\xi$: $\beta_A=\xi\beta_{\Cr}$\footnote{We shall still occasionally write $\beta_{\Ar}$ if it makes the notation more straightforward.}. With this, expression \eqref{eq:phi1} turns into 
\begin{equation}\label{eq:phi1a}\phi_{i}(X)=\beta_{\Cr}\left(\xi I_{Ai}+\sum_{j=1}^n \alpha_{ij} I_{Cj}\right).\end{equation}

\subsection{A generalized model with differentially effective control, variable transmissibility and prophylaxis}
We generalize the baseline model by allowing different strains to have different phenotypes by relaxing the model assumptions along the following lines:
\begin{itemize}
\item The treatment program does not ensure complete suppression of viral replication. That is to say, the treatment program fails with certain probability, which varies depending on the virus strain.
\item Virus strains have different transmissibility.
\item In addition to the treatment, we consider the effect of chemical or biological prophylaxis. While on prophylaxis, an individual acquires protection against the virus, the extent of which depends on the virus strain.
\end{itemize}

To account for different failure rates of treatment we divide the group of people on treatment into $n$ compartments $T_i$, where $i$ corresponds to the virus strain. Furthermore, we add a cohort of people receiving prophylaxis, denoted by $P$. While on prophylaxis, the individuals acquire variable protection against different virus strains denoted by $\psi_i\in[0,1]$ with $\psi_i=1$ corresponding to full protection. Thus we have the following model:

\begin{equation}\label{eq:sys2}\begin{aligned}
    \dot{I}_{Ai}={ }&\phi_{i}(X)S + (1-\psi_i)\phi_i(X)P -\gamma I_{Ai}-\mu I_{Ai}\\
    \dot{I}_{Ci}={ }&\gamma I_{Ai}+\zeta_i T_i - u_{\Tr} I_{Ci} - \mu I_{Ci}\\
    \dot{T_i}= { }& u_{\Tr} I_{Ci}-\zeta_i T_i -  \mu T_i\\
    \dot{S}= { }&\mu - u_{\Prm} S - \sum_{i=1}^n\phi_{i}(X)S +\delta P -\mu S\\
        \dot{P}= { }&u_{\Prm} S-\sum_{i=1}^n(1-\psi_i)\phi_i(X)P-\delta P-\mu P
\end{aligned}\end{equation}
where $\zeta_i\ge 0$ is the failure rate associated with the $i$th control, $\delta$ is the inverse duration of prophylaxis, and $u_{\Tr}$, resp., $u_{\Prm}$ are the rates at which people are administered to either treatment or prophylaxis. To account for variable transmissibility of different virus strains we define a set of transmissibility rates $\beta_{Ai}$ and $\beta_{Ci}$, $i=1,\dots,n$. Similarly to the baseline case, the transmissibility rates for the corresponding acute and chronic stages are assumed to be proportional, i.e., $\beta_{Ai}=\xi \beta_{Ci}$.  The forces of infection $\phi_i(X)$ are defined as
\begin{equation}\label{eq:phi2}\phi_{i}(X)=\beta_{Ci}\left(\xi I_{Ai}+\sum_{j=1}^n \alpha_{ij} I_{Cj}\right).\end{equation}
Note that setting either $\zeta_i=0$ or $\psi_i=0$ or $\beta_{Ci}=\beta_{\Cr}$ for all $i=1,\dots,n$, we obtain different variations of the baseline model.


\paragraph{Notation} We let $\mathbf{0}$, $\mathbf{1}$, and $\Er$ denote the matrices of zeros, ones, and the identity matrix\footnote{The use of notation $\Er$  instead of $\Ir$ for the identity matrix is common in German and Russian mathematical texts (Germ., {\bf E}inheitsmatrix). Here we use it to avoid confusing the notation $\Ir$ with the letter $I$ used for infected compartments.}. The sizes of the respective matrices are indicated as subscripts. A single subscript, e.g., as in $\Er_n$, denotes a square $[n\times n]$ matrix of respective type. Furthermore, $\Ir_A$ and $\Ir_{\Cr}$ denote the column vectors of respective variables and $\Ar$ denotes the matrix of $\alpha$'s:
$$\Ir_A=\begin{bmatrix}I_{A1}\\\vdots\\I_{An}\end{bmatrix},\enskip \Ir_{\Cr}=\begin{bmatrix}I_{C1}\\\vdots\\I_{Cn}\end{bmatrix},\enskip\mbox{and} \Ar=\begin{bmatrix}\alpha_{11}&\dots&\alpha_{1n}\\\vdots&\ddots&\vdots\\\alpha_{n1}&\dots&\alpha_{nn}\end{bmatrix}.$$
Note that $\Ar$ is a non-negative, column stochastic matrix, i.e., all its columns sum to 1.

All parameters and variables used in the model are listed in Table \ref{tab1}. Note that all quantities used are assumed to take on non-negative values and the index $i$ always runs from $1$ to $n$.

\begin{table}[hbt]
  \centering
  \caption{Model parameters}\label{tab1}\smallskip

\begin{tabular}{|c|c|p{12cm}|}
\hline
Variable&Range&Description\\
\hline
$I_{Ai}$ & $[0,1]$ & Fraction of acutely infected individuals infected by the virus of type $i$.\\[3pt]
$I_{Ci}$ & $[0,1]$ & Fraction of chronically infected individuals infected by the virus of type $i$.\\[3pt]
$S$ & $[0,1]$ & Fraction of susceptible individuals\\[3pt]
$T$ & $[0,1]$ & Fraction of patients involved in treatment\\[3pt]
$T_i$ & $[0,1]$ & Fraction of patients infected by the virus of type $i$ that are involved in treatment\\[3pt]
\hline
\hline
Parameter&Range&Description\\
\hline
$\gamma$ &  & Inverse duration of the acute phase\\[3pt]
$ u_{\Tr}$ & & Fraction of chronically infected that are enrolled into treatment (controlled parameter)\\[3pt]
$ u_{\Prm}$ & & Fraction of chronically infected that are enrolled into prophylaxis (controlled parameter)\\[3pt]
$\mu$     &  & Mortality rate\\[3pt]
$\alpha_{ij}$ & [0,1] & Fraction of type $i$ viruses in the viral population of an individual initially infected by the type $j$ virus.\\[3pt]
$\beta_A$, $\beta_{\Cr}$ &  & Transmissibility rates of acute and chronically infected individuals.\\[3pt]
$\xi$&&Proportionality coefficient of the transmissibility in acute and chronic stages\\[3pt]
$\zeta_i$&&Failure rate of treatment for individuals infected by the virus of type $i$\\[3pt]
$\delta$&&Failure rate of prophylaxis\\[3pt]

\hline
\end{tabular}
\end{table} 

\subsection{Structural analysis}
In this subsection we consider only the baseline model \eqref{eq:sys1} since the extended model has the same properties and can be readily analyzed along the same lines. 

\paragraph{Non-negativity of the solutions}
The equations \eqref{eq:sys1} can be written as 
\begin{equation}\label{eq:sys1a}
\frac{d}{dt}\begin{bmatrix}\Ir_A\\\Ir_{\Cr}\\T\\S\end{bmatrix}=
\begin{bmatrix}\beta_{\Cr}\left[\xi \Ir_A+\Ar\Ir_{\Cr}\right]S -(\gamma +\mu) \Ir_{A}\\\gamma \Ir_{A} - \mu \Ir_{\Cr}\\-\mu T\\\mu - \beta_{\Cr}\mathbf{1}_{[1\times n]}\left[\xi \Ir_A+\Ar\Ir_{\Cr}\right]S -\mu S\end{bmatrix}
+\begin{bmatrix}\mathbf{0}_{[n\times 1]}\\- \Ir_{\Cr}\\\mathbf{1}_{[1\times n]} \Ir_{\Cr}\\0\end{bmatrix}u_{\Tr}=\Psi(X)+\Psi^u(X) u_{\Tr}.
\end{equation}
The vector-valued functions $\Psi(X)$ and $\Psi^u(X)$ are {\em essentially non-negative}, i.e., for all $j=1,\dots, m$, $m=2n+2$, it holds that $\Psi_{j}(\tilde X)\ge 0$ (resp., $\Psi^u_{j}(\tilde X)\ge 0$) for any $\tilde X\in \R^m_{\ge 0}$ such that $\tilde X_j=0$ (see \cite{MMB:17} for details). This implies that solutions of \eqref{eq:sys1} are non-negative. That is to say, for any non-negative initial condition $X(0)=X_0\in\R^m_{\ge 0}$ and any non-negative control $u_{\Tr}$ the solution of \eqref{eq:sys1} belongs to $\R^m_{\ge 0}$ for all $t\ge 0$.

\paragraph{Boundedness of solutions}
Observe that the $m$-simplex $\Delta_m$, formed as the convex hull of $m$ unit vectors $\mathbf{e}_j$, $j=1,\dots,m$, is invariant with respect to \eqref{eq:sys1}: $$X(0)\in \Delta_m \Rightarrow X(t)\in \Delta_m,$$
where $\Delta_m=\{X\in \R^m_{\ge 0}|\sum_{j=1}^m X_j=1\}$. This result follows immediately from the fact that the states $X_i$ represent the fractions of the respective groups within the total population and hence sum to 1.

\section{Local analysis at a disease-free equilibrium}

Below, we will compute the basic reproduction number for two considered models and present a number of related results. To distinguish between the basic reproduction numbers related to different models we will add a superscript denoting the particular model: $\alpha$ for the baseline model \eqref{eq:sys1} and $\beta$ for the extended model \eqref{eq:sys2}.

\subsection{Basic reproduction number for the baseline model}
The system \eqref{eq:sys1} has a unique disease-free equilibrium (DFE) $X_{DFE}=[0,\dots,0,1]$. To analyze the stability property of the system \eqref{eq:sys1} at the DFE we compute the controlled basic reproduction number $R_0$ using the classical next-generation method \cite{Driessche:02} (see \cite{JTB:19} for an extension of the method that takes into account the action of a control). 

\begin{thm}\label{thm_R01}
For any choice of parameters $\alpha_{ij}\ge 0$ such that $\sum_i\alpha_{ij}=1$ and $\alpha_{ii}\neq 0$ for all $i,j=1,\dots,n$, the controlled basic reproduction number of the system \eqref{eq:sys1} is given by
\begin{equation}\label{eq:R0}R^\alpha_0(u_{\Tr})=\beta_{\Cr}\frac{\xi( u_{\Tr}+\mu)+\gamma}{(\gamma+\mu)( u_{\Tr}+\mu)}=\frac{\xi \beta_{\Cr}}{\gamma+\mu}+\frac{\beta_{\Cr}\gamma}{(\gamma+\mu)( u_{\Tr}+\mu)}.\end{equation}
\end{thm}
\begin{pf}
See Appendix B.
\end{pf}
Note that the $\alpha_{ij}$ values do not affect the basic reproduction number, which makes sense in this context as mutation from one strain into another does not imply any change in a relevant phenotype such as contagiousness or resistance to therapy. In this context, a different strain simply carries a distinct mutation (or pattern of mutations) that makes it identifiable from other strains. However, understanding the distribution of strains with the same phenotype is an important aspect of molecular epidemiology, which is dependent on the specific  $\alpha_{ij}$ values. This relationship between within-host mutations and endemic equilibrium of infection types is discussed below.

\paragraph{Sensitivity analysis} When devising an intervention strategy, the main question to be answered is whether the proposed treatment or prophylaxis scheme is capable of eliminating the infection, i.e., driving the basic reproduction number below $1$. To address this issue we introduce the sensitivity parameter(s) $R_1$ that quantify the efficiency of sufficiently small controls in reducing the value of $R_0$, \cite{JTB:19}. In particular, the controlled basic reproduction number $R_0^\alpha(u_{\Tr})$ is expanded as
\begin{equation}\label{eq:R0al}R^\alpha_0(u_{\Tr})=R^\alpha_0+ R^\alpha_1 u_{\Tr} +\mathcal{O}(u_{\Tr}^2),\end{equation}
where $R^\alpha_0=R_0^\alpha(0)=\beta_{\Cr}\frac{\gamma +\xi\mu }{\mu \left(\gamma +\mu \right)}$ and $R^\alpha_1=-\frac{\beta_{\Cr}\gamma}{\mu ^2\left(\gamma +\mu \right)}$. 
Before proceeding with the further analysis, we define the notion of efficiency of a control.
\begin{defn}\label{def:global-eff}Let the uncontrolled basic reproduction number be larger than 1, i.e., $R_0(0)>1$. A control $u$ is said to be 
\begin{enumerate}
\item {\em locally efficient} if the respective sensitivity parameter is negative, i.e., $R^u_1<0$;
\item {\em (globally) efficient} if there exists a non-negative value $u^*$ such that $R_0(u^*)=1$.
\end{enumerate}
Furthermore, we say that a control is {\em unconditionally} locally (globally) efficient if 1.\ (2.) holds for all admissible values of parameters. Otherwise the control is said to be {\em conditionally} efficient.
\end{defn}

We can immediately observe that $u_{\Tr}$ is unconditionally locally efficient. However, an unconditionally locally efficient control may fail to reach the stated goal of eliminating the infection, i.e., reducing $R_0$ below 1. The following result illustrates that. 

\begin{lem}\label{lem:efficiency} The control $u_{\Tr}$ is  efficient if $\beta_{\Cr}$ satisfies 
\begin{equation}\label{eq:cond-betaC}
\xi\beta_{\Cr}  < \gamma+\mu.
\end{equation}
\end{lem}
\begin{pf}This result can be easily checked by observing the expression for $R^\alpha(u_{\Tr})$ in \eqref{eq:R0} and noting that the second summand vanishes as $u_{\Tr}$ tends to infinity.\qed\end{pf}
\begin{rem}
Note that the condition \eqref{eq:cond-betaC} can be alternatively rewritten as $\beta_A\theta_A<1$, where $\theta_{\Ar}=1/(\gamma+\mu)$ denotes the average duration of the acute stage.
\end{rem}

The result of Lemma \ref{lem:efficiency} implies that the control $u_{\Tr}$ is only {\em conditionally} globally efficient. That is, it can be used to completely eliminate the infection only if the transmissibility $\beta_{\Cr}$ satisfies \eqref{eq:cond-betaC}.

\subsection{Basic reproduction number for the extended model}

In contrast to the baseline case, the disease free equilibrium for the modified model \eqref{eq:sys2} is shifted due to the action of the control $u_{\Prm}$. So, we have
\begin{equation}\label{eq:DFE2}X_{DFE}=[\mathbf{0}_{1\times n},\, \mathbf{0}_{1\times n},\, \mathbf{0}_{1\times n},\, P_{DFE},\, S_{DFE}],\end{equation} 
where $S_{DFE}(u_{\Prm})=\frac{\delta+\mu}{\delta+\mu+u_{\Prm}}$ and $P_{DFE}(u_{\Prm})=1-S_{DFE}(u_{\Prm})=\frac{u_{\Prm}}{\delta+\mu+u_{\Prm}}$. Local stability of the DFE \eqref{eq:DFE2} is determined by $R^\beta_0(u_{\Tr},u_{\Prm})$. Before we proceed with the analysis, we note that the results to follow will be formulated using matrix notation. In particular, we will write $\Br_{\Cr}=\mathrm{diag}(\beta_{\Cr 1},\,\dots,\, \beta_{\Cr n})$, $\Psi=\mathrm{diag}(\psi_1,\,\dots,\, \psi_n)$, and $Z=\mathrm{diag}(\zeta_1,\,\dots,\, \zeta_n)$ to denote the diagonal matrices of transmissibility rates, protection factors and treatment failure rates.

\begin{thm}\label{thm_R02}
The controlled basic reproduction number of the system \eqref{eq:sys2} is given by
\begin{equation}\label{eq:R02}R^\beta_0(u_{\Tr},u_{\Prm})=\frac{\bar\beta_{\Cr}(\gamma+\xi\mu)}{(\gamma+\mu)\mu}\rho\left(Q(u_{\Prm})N(u_{\Tr})\right),\end{equation}
where $\bar\beta_{\Cr}=\max_{i}\beta_{Ci}$, $\bar \Br_{\Cr}=\bar\beta_{\Cr}^{-1} \Br_{\Cr}$, $Q(u_{\Prm})=\bar \Br_{\Cr}\left[\Er_n-P_{DFE}(u_{\Prm})\Psi\right]$, $N(u_{\Tr})=\frac{1}{\gamma+\xi\mu}\left[\xi\mu\Er_n+\gamma \Ar\Delta(u_{\Tr})\right]$, and $\Delta(u_{\Tr})=(\Zr +(\mu+u_{\Tr}) \Er_n)^{-1}(\Zr+\mu\Er_n)$.
\end{thm}
\begin{pf}
See Appendix B.
\end{pf}

Note that the basic reproduction number of the extended system is a product of two terms: the first one closely resembles $R^\alpha_0$ as in \eqref{eq:R0al}, while the second term is a spectral radius of the product of two matrices, where the first one depends only on $u_{\Prm}$ and the second one depends only on $u_{\Tr}$.  

Before we proceed with the analysis, we formulate an important result on stochastic matrices.

\begin{lem}\label{lem:A(uT)}
Let $\Sigma$ be a non-negative, column stochastic matrix. Then for any $\alpha\in[0,1]$, the convex combination $\Sigma_\alpha=\alpha\Er+(1-\alpha)\Sigma$ is a column stochastic matrix as well.  Furthermore, the left and right dominant eigenvectors of $\Sigma$ coincide with those of $\Sigma_\alpha$.
\end{lem}
The above result implies that $N(0)=\frac{1}{\gamma+\xi\mu}\left[\xi\mu\Er_n+\gamma \Ar\right]$ is a column stochastic matrix, whose left and right dominant vectors coincide with those of $\Ar$. We will thus write $N(0)=\bar\Ar$.

\paragraph{Sensitivity analysis} We begin this paragraph by writing down an expansion of $R^\beta_0(u_{\Tr},u_{\Prm})$.

\begin{thm}\label{thm:R0-beta} Let $A$ be irreducible and let $w_0$ and $v_0$ be the right and the left dominant eigenvectors of $Q(0)N(0)=\bar \Br_{\Cr}\bar \Ar$, corresponding to $\rho\left(\bar \Br_{\Cr}\bar \Ar\right)$ and normalized such that $w^\top_0 v_0=1$. The controlled basic reproduction number $R^\beta_0(u_{\Tr},u_{\Prm})$ can be written as
\begin{equation}\label{eq:R0be}R^\beta_0(u_{\Tr},u_{\Prm})=R^\beta_0+ R^\beta_{1,\Tr} u_{\Tr}+ R^\beta_{1,\Prm} u_{\Prm} +\mathcal{O}(\|(u_{\Tr},u_{\Prm})\|^2),\end{equation}
where $R^\beta_0=\frac{\bar\beta_{\Cr}(\gamma+\xi\mu)}{(\gamma+\mu)\mu}\rho\!\left(\bar \Br_{\Cr}\bar \Ar\right)$, $R^\beta_{1,\Tr}=-w_0^\top\left[R_0^\beta \Er_n - \frac{\xi}{(\gamma+\mu)} \Br_{\Cr}\right](\Zr+\mu\Er_n)^{-1}v_0$, and $R^\beta_{1,\Prm}=-R_0^\beta \frac{1}{(\delta+\mu)} w_0^\top \Psi v_0$.
\end{thm}
\begin{pf}See Appendix B.\end{pf}
This result has a number of important consequences as formulated below. We first consider a slightly simplified setup.
Let there be no variability in transmission rates, i.e., $\Br_{\Cr}=\beta_{\Cr}\Er_n$. Then the vectors $w_0$ and $v_0$ coincide with those of $A$. Furthermore, we have $w_0=\mathbf{1}_{[n\times 1]}$. The respective coefficients turn into $R^\beta_0=R^\alpha_0$, $R^\beta_{1,\Tr}=-R^\beta_0\frac{\gamma}{\gamma+\xi\mu} w_0^\top(\Zr+\mu\Er_n)^{-1}v_0$, and $R^\beta_{1,\Prm}=-R_0^\beta \frac{1}{(\delta+\mu)} w_0^\top \Psi v_0$. That is, we can write
\begin{equation*}R^\beta_0(u_{\Tr},u_{\Prm})=R^\beta_0\left(1-\frac{\gamma}{\gamma+\xi\mu} w_0^\top(\Zr+\mu\Er_n)^{-1}v_0 \cdot u_{\Tr}  - \frac{1}{(\delta+\mu)} w_0^\top \Psi v_0 \cdot u_{\Prm}\right) +\mathcal{O}(\|(u_{\Tr},u_{\Prm})\|^2).\end{equation*}
Obviously, both controls are unconditionally locally efficient. We can also observe that the control $u_{\Tr}$ is locally more efficient than $u_{\Prm}$ if it holds that 
\begin{equation}\label{eq:ineq1}\frac{\gamma}{\gamma+\xi\mu} w_0^\top(\Zr+\mu\Er_n)^{-1}v_0 > \frac{1}{(\delta+\mu)} w_0^\top \Psi v_0.\end{equation}
Obviously, we have that $u_{\Prm}$ is locally more efficient if the opposite holds true. The inequality \eqref{eq:ineq1} can be interpreted as follows. Note that $\tau_i=1/(\zeta_i+\mu)$ and $\pi=1/(\delta+\mu)$ are the average duration of being either on treatment or on prophylaxis and recall that $w_0^\top=[1,\dots, 1]$. Then we can write \eqref{eq:ineq1} as 
\begin{equation*}\sum_i\frac{\gamma}{\gamma+\xi\mu}\tau_i v_{0i} > \sum_i \psi_i\pi v_{0i}.\end{equation*}
Here, the factor  $\gamma/(\gamma+\xi\mu)$ is interpreted as the degree of protection given by the treatment. Note that this number decreases with increasing $\xi$, i.e., when the acute stage is much more contagious compared to the chronic stage. If $\xi=1$, the fraction $\gamma/(\gamma+\mu)$ merely corresponds to the fraction of people that survive to the chronic stage. Note that this interpretation has to do with the fact that we assume the acute stage is short enough that people will not start treatment conditional on being in the chronic stage of infection and therefore the acute to chronic stage contagiousness of infection is a major determinant of the efficacy of therapy as a population-level control. This assumption is reasonable for diseases like HIV, but may need to be revisited for application to other diseases.  Next, we note that the components of the vector $v_0$ are proportional to the stationary distribution of different strains of the virus in the baseline model (see Sec.\ \ref{sec:EE} for more details on that). Thus, we can interpret the sensitivity parameters $R_{1,\Tr}^\beta$ and $R_{1,\Prm}^\beta$ as a sum of products {\em average duration of the medical intervention} $\times$ {\em protection conferred by the intervention} taken with the weights corresponding to the stationary distribution of the virus strains. 

Following the same line, one can attempt to compare the efficiency of two controls in the general case. To start with, we write \eqref{eq:R0be} as 
$$R^\beta_0(u_{\Tr},u_{\Prm})=R^\beta_0\left(1-w_0^\top\left[\Er_n - \frac{\xi}{(\gamma+\mu)R^\beta_0} \Br_{\Cr}\right](\Zr+\mu\Er_n)^{-1}v_0 u_{\Tr}- \frac{1}{(\delta+\mu)} w_0^\top \Psi v_0 u_{\Prm}\right) +\mathcal{O}(\|(u_{\Tr},u_{\Prm})\|^2)$$
As above, we say that $u_{\Tr}$ is locally more efficient than $u_{\Prm}$ if 
\begin{equation}\label{eq:ineq2}\sum_i \left[1 - \frac{\beta_{\!\Ar i}\theta_{\Ar}}{R^\beta_0}\right]\tau_i w_{0i} v_{0i} > \sum_i \psi_i \pi w_{0i} v_{0i}.\end{equation}
Similarly to the previous case, we interpret the expression in front of $\tau_i$ as the degree of protection given by the treatment to those infected with the $i$-type virus. Note that a sufficient condition for this expression to be positive is $\beta_{Ai}\theta_A<1$. In contrast to the previous case, the components $w_{0i} v_{0i}$ do not have that clear interpretation. However, their behavior is pretty close to that of $v_{0i}$ as the numerical simulation presented in Fig.\ \ref{fig:w-v} illustrates. 

Finally, we observe that for sufficiently large controls $u_{\Tr}$ and $u_{\Prm}$ we have
\begin{equation*}\lim_{u_{\Tr}\rightarrow \infty\atop u_{\Prm}\rightarrow \infty}R^\beta_0(u_{\Tr},u_{\Prm})
=\frac{\xi}{(\gamma+\mu)}\rho\left(\Br_{\Cr}\left[\Er_n-\Psi\right]\right)=\frac{\xi}{(\gamma+\mu)}\max_i(\beta_{\Cr,i}(1-\psi_i)),\end{equation*}
which yields the result that agrees with the result of Lemma \ref{lem:efficiency}.
\begin{lem}
The controls $u_{\Tr}$ and $u_{\Prm}$ are jointly globally efficient if
$$\xi\max_i(\beta_{\Cr,i}(1-\psi_i))<\gamma+\mu.$$
\end{lem}
\section{Endemic equilibrium} \label{sec:EE}

In contrast to the unique disease-free equilibrium, there can be one, many (a continuum), and no endemic equilibria at all. Which case realizes in our system depends on the value of the basic reproduction number and on the structure of the matrix $\Ar$ as will be shown below. For the general case, the endemic equilibrium can be computed using a rather involved semi-analytic procedure and offers a little insight into the structure of the respective equilibrium. Therefore, we restrict ourselves to the baseline model and a particular extension thereof. The general model is considered in Section \ref{sec:num} that is devoted to the numerical simulations.

\subsection{A baseline model}
We begin by stating a general result on the endemic equilibrium.

\begin{thm}\label{thm2}
Let $\Ar$ be an irreducible non-negative column stochastic matrix such that all diagonal elements are non-zero. Then the endemic equilibrium for the system \eqref{eq:sys1} exists and is unique if $R_0>1$. Let, furthermore, $v^\top=[v_1,\dots,v_n]$ be the right normalized eigenvector of $\Ar$ corresponding to the dominant eigenvalue of $\Ar$ and satisfying $\sum_{i=1}^n v_i=1$. The components of the endemic equilibrium state are given by
\begin{equation}\label{eq:eq-sol}\begin{aligned}
I^*_{Ai}={ }&\frac{\mu}{(\gamma+\mu)}\left(1-\frac{1}{R_0}\right)v_i,\qquad\quad I^*_{Ci}=\frac{\gamma \mu}{(\gamma+\mu)(u_{\Tr}+\mu)}\left(1-\frac{1}{R_0}\right)v_i,\\
T^*={ }& \frac{\gamma  u}{(\gamma+\mu)(u_{\Tr}+\mu)}\left(1-\frac{1}{R_0}\right), \enskip S^*= \frac{1}{R_0}.
\end{aligned}\end{equation}
\end{thm}
\begin{pf}
See Appendix B.
\end{pf}

Note that the only additional property of the matrix $\Ar$ that is involved in this theorem is that $\Ar$ is {\em irreducible}. For the definition of irreducibility and further details see Appendix A. 

The obtained result can be used to compute a number of derived quantities. For instance, we have that the total prevalence at the endemic equilibrium is equal to 
$$\Pi=1-S^*=\frac{R_0-1}{R_0}$$
and the the ratio of transmissions through acutely infected to the transmission through chronically infected is given by
\begin{equation}\label{eq:rAC}
    r_{AC}= \frac{\xi\beta_{\Cr} \sum_{i=1}^n I_{A,i}}{\beta_{\Cr} \sum_{i=1}^n I_{C,i}}=\frac{\xi\beta_{\Cr} i_{A\Sigma}}{\beta_{\Cr} i_{C\Sigma}}=\frac{\xi\beta_{\Cr} (u_{\Tr}+\mu)}{\beta_{\Cr} \gamma}.
\end{equation}
Using the statistical estimations of these two parameters one can recover $\xi$ and $\beta_{\Cr}$.

Before we proceed to the next result we recall that $\alpha_{ij}$ can be interpreted as the probability of catching a virus of type $i$ through the contact with an individual initially infected by the virus of type $j$. So, we can make the following observation.

\begin{lem}
At the endemic equilibrium, the probability of encountering a chronically infected in the $i$th category is equal to the probability of catching the type $i$ virus through the contact with a randomly chosen chronically infected individual:
$$I^*_{Ci}=\sum_{j=1}^n  \alpha_{ij} I^*_{Cj}.$$
\end{lem}
\begin{pf}
Using the expression for $I^*_{Ci}$, we can write
$$\sum_{j=1}^n  \alpha_{ij} I^*_{Cj}=\frac{\gamma \mu}{(\gamma+\mu)(u_{\Tr}+\mu)}\left(1-\frac{1}{R_0}\right)\sum_{j=1}^n  \alpha_{ij} v_j.$$
However, since $v$ is the stationary eigenvector of $\Ar$, it holds that $\sum_{j=1}^n  \alpha_{ij} v_j=v_i$, whence the result follows.\qed
\end{pf}

If the matrix $\Ar$ is reducible, the results of Theorem \ref{thm2} do not apply any longer. However, we can formulate a weaker version of the theorem. First, we note that a reducible matrix can be transformed to the normal form by means of a properly chosen permutation matrix:
\begin{equation}\label{eq:normf}
\tilde \Ar = PAP^\top = \begin{bmatrix}
\tilde \Ar_1&\ast&\dots&\ast\\
\mathbf{0}&\tilde \Ar_2&\dots&\ast\\
\vdots&&\ddots&\ast\\
\mathbf{0}&\mathbf{0}&\dots&\tilde \Ar_k
\end{bmatrix},\end{equation}
where $\Ar_i$, $i=1,\dots,k$ are irreducible matrices and asterisks denote arbitrary non-negative matrices.
\begin{thm}\label{thm3} Let $\Ar$ be a reducible non-negative matrix with non-zero diagonal elements such that it can be transformed into the normal form \eqref{eq:normf} by an appropriate simultaneous permutation of rows and columns. Then $\Ar$ has at most $k$ unit eigenvalues. Furthermore, let $\mathbf{v}=\{v_1,\dots,v_q\}$ be the set of normalized eigenvectors corresponding to the unit eigenvalues, $q\le k$. Then the set of endemic equilibria is defined as follows:
\begin{equation}\label{eq:eq-sol2}\begin{aligned}
I^*_{Ai}={ }&\frac{\mu}{(\gamma+\mu)}\left(1-\frac{1}{R_0}\right)\bar v_i,\qquad\quad I^*_{Ci}=\frac{\gamma \mu}{(\gamma+\mu)(u_{\Tr}+\mu)}\left(1-\frac{1}{R_0}\right)\bar v_i,\\
T^*={ }& \frac{\gamma  u}{(\gamma+\mu)(u_{\Tr}+\mu)}\left(1-\frac{1}{R_0}\right), \enskip S^*= \frac{1}{R_0},
\end{aligned}\end{equation}
where the vector $\bar v$ belongs to the linear hull of vectors from $\mathbf{v}$: $\bar v\in \mathrm{Span}(\mathbf{v})$.
\end{thm}
Theorem \ref{thm3} implies that the set of endemic equilibria can form a linear subspace of the system's state space. The case when the matrix $\Ar$ is reducible corresponds to the situation when there are some particular groups of virus strains, say, two groups $G_1$ and $G_2$. Reducibility implies that the mutations between these groups are either not possible at all, $G_1 \nleftrightarrow G_2$ or possible only in one direction, $G_1 \leftarrow G_2$, but $G_1 \nrightarrow G_2$ (or vice versa). Such a setup allows for considering directed patterns of viral evolution. However, this question is beyond the scope of this paper and will be addressed in our future work.
\paragraph{Structure of the matrix $A$: uniform within host mutations}

An important observation that follows from the preceding analysis  is that one cannot unambiguously determine all $n^2$ parameters $\alpha_{ij}$ from the observations made at the endemic equilibrium. The reason for that is that the equilibrium values depend on the $n$ components of the vector $v$ (see \eqref{eq:eq-sol}) of which only $n-1$ values are independent. We thus restrict ourselves to considering one particular structure of the matrix $A$ that can be formulated in terms of only $n$ parameters. More complex structures are possible and can be treated using the same results. In particular, Theorem \ref{thm:L} offers a convenient tool for computing the respective dominant eigenvector.

Assume that during the chronic infection stage the viral population of the individual, initially infected with the type $i$ virus contains the fraction $\alpha_{i}$ of the original virus while the remaining strains of the virus are distributed uniformly. This means that the matrix $\Ar$ has the following form:
\begin{equation}\label{eq:A-ex1}\Ar=\begin{bmatrix}
\alpha_{1}&\frac{1-\alpha_{2}}{n-1}&\dots&\frac{1-\alpha_{n}}{n-1}\\
\frac{1-\alpha_{1}}{n-1}&\alpha_{2}&&\frac{1-\alpha_{n}}{n-1}\\
\vdots&\vdots&\ddots&\vdots\\
\frac{1-\alpha_{1}}{n-1}&\frac{1-\alpha_{2}}{n-1}&\dots&\alpha_{n}\\
\end{bmatrix}.\end{equation}
The matrix \eqref{eq:A-ex1} is positive hence, the Perron-Frobenius theorem applies. There is a unique dominant eigenvalue that is equal to 1, and the components of the dominant eigenvector have the following form:
$$v_i=\frac{\prod_{j\neq i}(1-\alpha_{j})}{\sum_{i=1}^n\prod_{j\neq i}(1-\alpha_{j})}.$$
The respective expressions for the system states at endemic equilibrium are pretty bulky. However, we can compute the relative ratios of infected in different groups, which turn out to have a simple form:
$$r_{ij}=\frac{I_{Ai}+I_{Ci}}{I_{Aj}+I_{Cj}}=\frac{v_i}{v_j}=\frac{1-\alpha_j}{1-\alpha_i}.$$
Note that the condition $\sum_i v_i=1$ implies that there are only $(n-1)$ independent equations. Thus, one parameter $\alpha_i$ can be set to an arbitrary value within the range $[0,1]$. Let, for instance, $\alpha_n$ be used as a free parameter. In this case, all remaining probabilities can be expressed in terms of $\alpha_n$ and $v_i$:
$$a_j=1-(1-\alpha_n)\frac{v_n}{v_j},\quad j=1,\dots,n-1.$$
In the following we will consider a slightly more realistic scenario in which all viruses are ordered according to their genetic similarity and any virus can mutate only to its ``neighbors''. The respective matrix $A$ has the following form:
\begin{equation}\label{eq:A-ex2}\Ar=\begin{bmatrix}
\alpha_{1}&\frac{1-\alpha_{2}}{2}&\dots&0\\
1-\alpha_{1}&\alpha_{2}&&0\\
0&\frac{1-\alpha_{2}}{2}&\dots&0\\
\vdots&\vdots&\ddots&\vdots\\
0&0&\dots&\alpha_{n}\\
\end{bmatrix}.\end{equation}
Quite remarkably, the respective expressions do not change that much compared to the previous case. Setting $v_i$ and assuming that $\alpha_n$ can be freely chosen we get  
$$\alpha_1=1-(1-\alpha_n)\frac{v_n}{v_1},\enskip \alpha_j= 1-2(1-\alpha_n)\frac{v_n}{v_{j}},\quad j=2,\dots,n-1.
$$
\subsection{Variable transmissibility}

In this subsection we consider an extension of the baseline model in which different strains are assumed to have different transmissibility rates. This implies that we consider the model \eqref{eq:sys2} with $u_P=0$ and $\zeta_i=0$. The respective set of DEs is written (in vector notation) as
\begin{equation}\label{eq:sys2a}\begin{aligned}
    \dot{I}_A={ }&B_{\Cr}\left(\xi I_{A}+A I_{\Cr}\right) S -\gamma I_{A}-\mu I_{A}\\
    \dot{I}_{\Cr}={ }&\gamma I_{A} - u_{T} I_{\Cr} - \mu I_{\Cr}\\
    \dot{T}= { }& u_{T} \sum I_{\Cr} -  \mu T\\
    \dot{S}= { }&\mu - \sum B_{\Cr}\left(\xi I_{A}+A I_{\Cr}\right) S  -\mu S.
\end{aligned}\end{equation}
For this case, the results of Theorems \ref{thm_R02}  and \ref{thm2} can be restated as follows:
\begin{thm}\label{thm2a}
The basic reproduction number $R^\beta_0(u_{\Tr})$ for the system \eqref{eq:sys2a} is given by
\begin{equation}R^\beta_0(u_{\Tr})=R^\beta_0(u_{\Tr},0)=\frac{\bar\beta_{\Cr}(\xi(\mu+u_{\Tr})+\gamma)}{(\gamma+\mu)(\mu+u_{\Tr})}\rho\left(\bar B_{\Cr} \bar{\bar A}\right),\end{equation}
where $\bar{\bar A}=\frac{1}{\xi(\mu+u_{\Tr})+\gamma}\left[\xi(\mu+u_{\Tr})\Er_n+\gamma \Ar\right]$. 
Let, furthermore, $\Ar$ be an irreducible non-negative column stochastic matrix such that all diagonal elements are non-zero. Let, furthermore, $v^\top=[v_1,\dots,v_n]$ be the right normalized dominant eigenvector of $\bar \Br_{\Cr}\bar{\bar \Ar}$ satisfying $\sum_{i=1}^n v_i=1$. Then the endemic equilibrium exists and unique if $R_0(u_{\Tr})>1$ and the respective components of the endemic equilibrium state are given by
\begin{equation}\label{eq:eq-sol}\begin{aligned}
I^*_{Ai}={ }&\frac{\mu}{(\gamma+\mu)}\left(1-S^*\right)v_i,\qquad\quad I^*_{Ci}=\frac{\gamma \mu}{(\gamma+\mu)(u_{\Tr}+\mu)}\left(1-S^*\right)v_i,\\
T^*={ }& \frac{\gamma  u}{(\gamma+\mu)(u_{\Tr}+\mu)}\left(1-S^*\right), \enskip S^*= \frac{1}{R_0(u_{\Tr})}.
\end{aligned}\end{equation}
\end{thm}
\begin{pf}
The expression for $R_0(u_{\Tr})$ is obtained from \eqref{eq:R02} after some algebraic manipulations. The rest of the proof closely follows the proof of Thm.\ \ref{thm2}.\hfill\qed
\end{pf}

We note that $\bar{\bar A}$ is a stochastic matrix and has the same right and left dominant eigenvectors as $A$ due to Lemma \ref{lem:A(uT)}. The product $\bar B_{\Cr}\bar{\bar A}$ corresponds to multiplying the rows of $\bar{\bar A}$ with the respective $\bar\beta_{Ci}$. For this expression one cannot find the dominant eigenvector analytically, so we consider some numerical examples to illustrate the influence of variable transmissibility on the distribution of different virus strains.

\begin{exmp}\label{exmp:1} Consider a model with 4 virus strains. We take the following values of the parameters: $\mu=0.025$, $\gamma=3$ (i.e., the acute phase takes about 4 months); $\xi=5$ (during the acute phase an individual is 5 times more contagious as in the chronic one); $u_{\Tr}=0.4$ (it takes 2.5 years on average till the treatment begins). The transmissibility rate $\beta_{\Cr}$ was chosen such that $R_0(u_{\Tr}=0.4)\approx 1.2$: $\beta_{\Cr}=0.25$. 
The matrix $A$ is assumed to have the form \eqref{eq:A-ex2} and the probabilities $\alpha_i$ are chosen in the way that the endemic distribution of the different virus strains for the baseline model satisfies $v_j/v_{j+1}=3$. That is, the endemic frequencies are $[v_1,\,v_2,\,v_3,\,v_4]= [0.675,\,0.225,\,0.075,\,0.025]$. Finally, we set $\alpha_4=0.25$. The remaining probabilities are computed to be $\alpha_1\approx 0.97$, $\alpha_2\approx 0.83$, and $\alpha_3= 0.5$.

To see how varying transmissibility influences the endemic distribution we fix the transmissibilities of the first 3 strains to be equal to $\beta_{\Cr}$, while the transmissibility of the fourth strain is $\beta_{C4}=a\beta_{\Cr}$, where $a$ changes from $0.7$ to $2$. The resulting endemic frequencies are shown in Fig.\ \ref{fig:freq}a. Note that at $a=1$ the endemic distribution coincides with the baseline one.
\begin{figure}[tbh]
\includegraphics[width=0.5\textwidth]{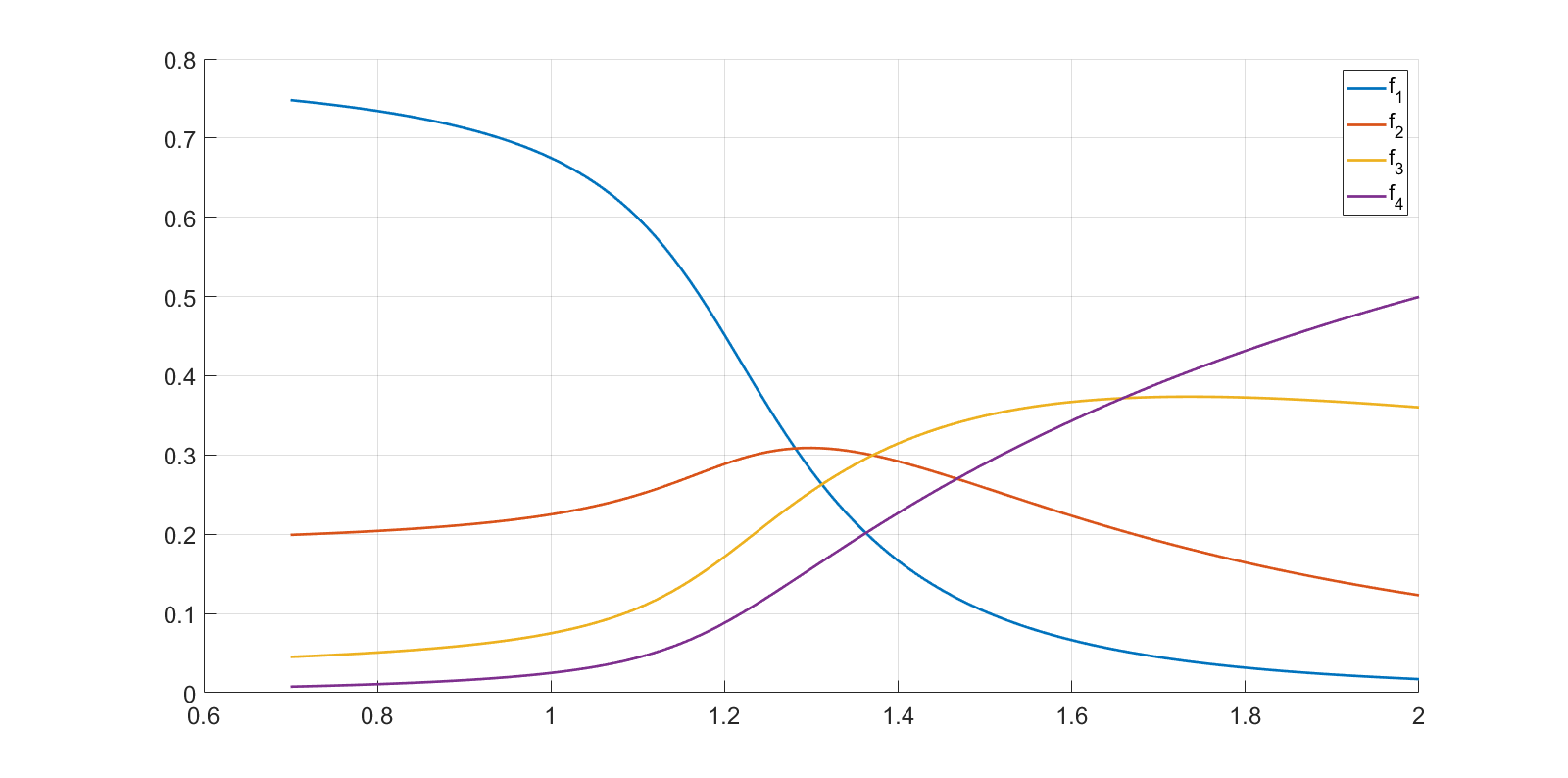}\includegraphics[width=0.5\textwidth]{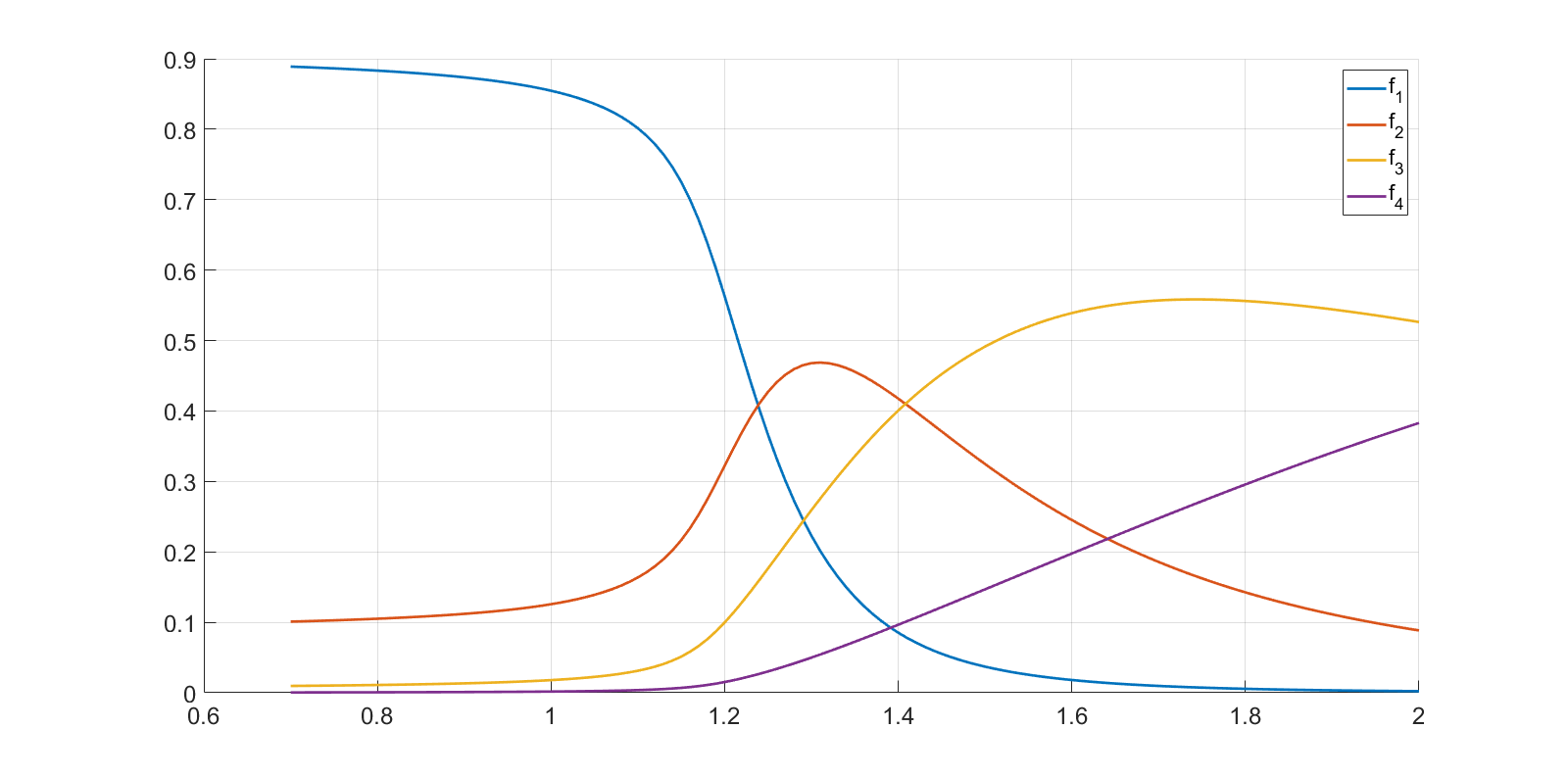}
\hspace*{3cm}a)\hspace*{8cm}b)
\caption{The endemic distribution of acutely infected ($I_{Ai}$) for different values of the transmissibility rate of the $4$th strain, parametrized with $a$: $\beta_{\Cr,4}=a\beta_{\Cr}$. The values at $a=1$ correspond to the baseline case (all transmissibility rates are equal). Subfigures a) and b) correspond to different values of mutation probabilities $\alpha_i$.}\label{fig:freq}
\end{figure}

\begin{figure}[tbh]
\includegraphics[width=0.5\textwidth]{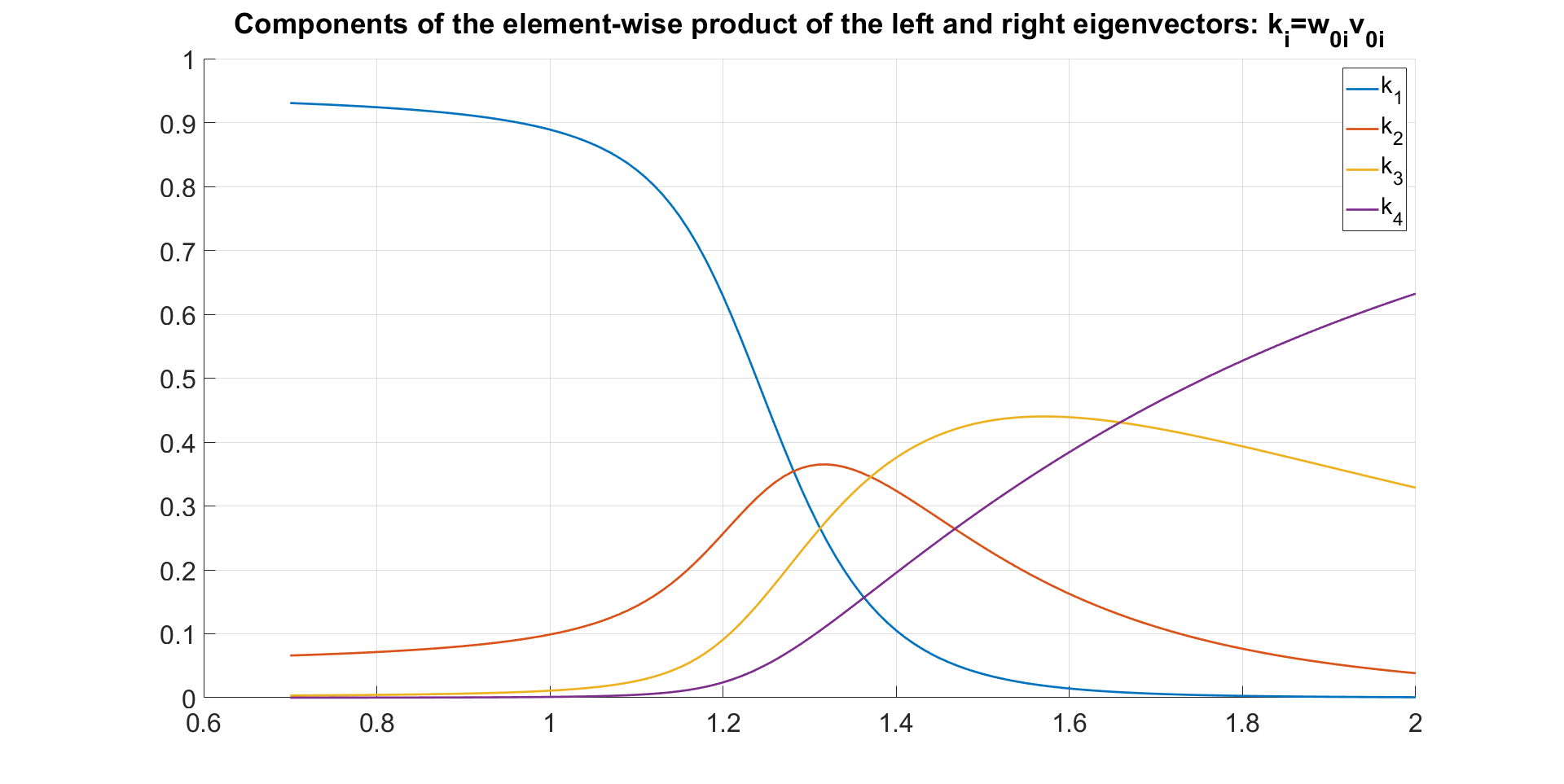}\includegraphics[width=0.5\textwidth]{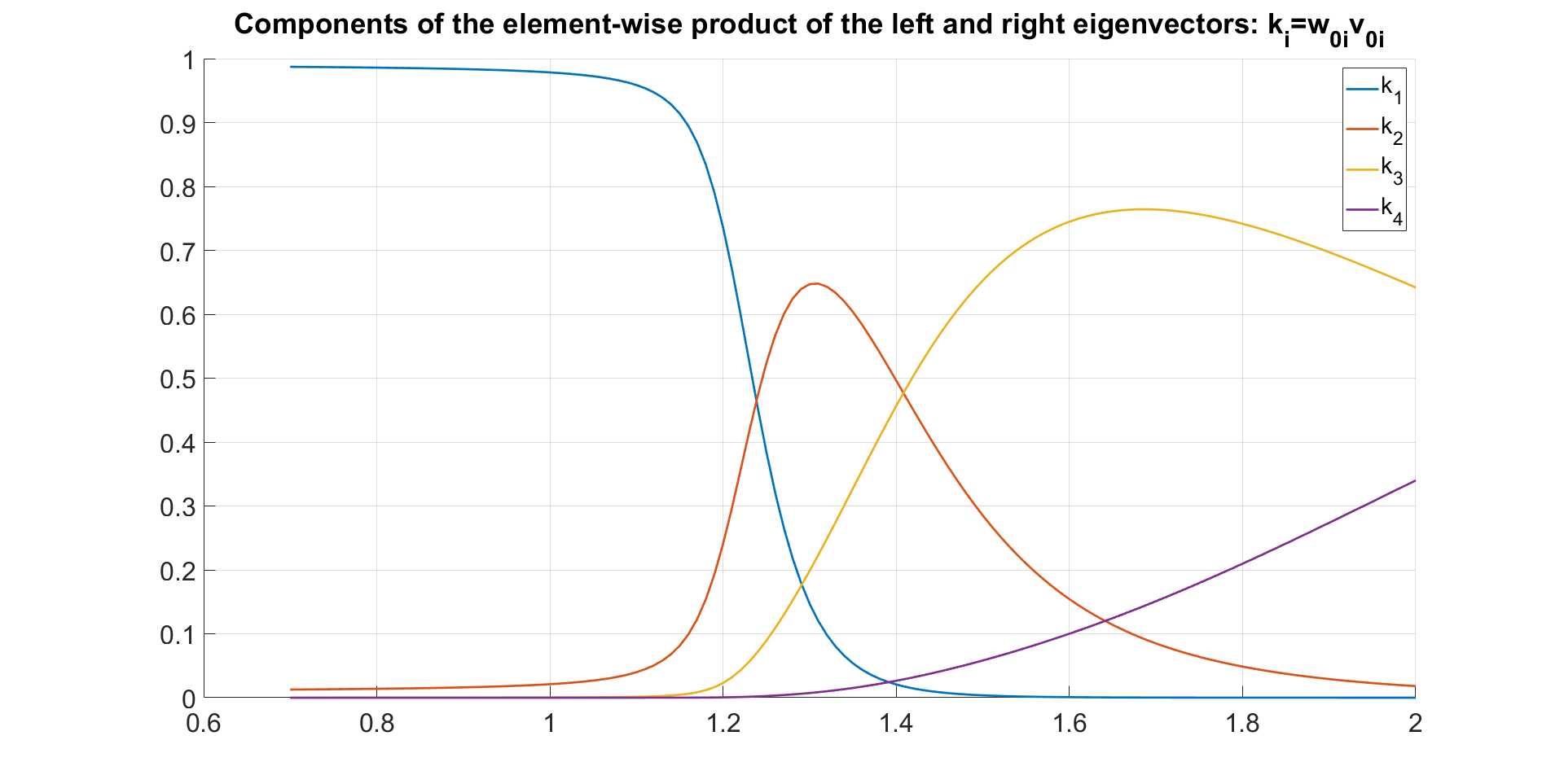}
\hspace*{3cm}a)\hspace*{8cm}b)
\caption{The change of the weights $w_{0i}v_{0i}$ that enter the sensitivity terms $R^\beta_{1,T}$ and $R^\beta_{1,P}$ with the parameter $a$.}\label{fig:w-v}
\end{figure}

To make a comparison, we consider a different set of parameters $\alpha_i$. In this case, we assume that these parameters are chosen such that the endemic distribution in the baseline case satisfies $v_i/v_{i+1}=7$. The respective values are $\alpha_1 = 0.9985$, $\alpha_2 = 0.9796$, $\alpha_3 = 0.8571$ and $\alpha_4 = 0.25$ (the last parameter was chosen to be equal to $\alpha_4$ in the previous case). The trajectories of the endemic frequencies are shown in Fig.\ \ref{fig:freq}b. It is interesting to observe that variation in transmissibility of one strain leads to substantial variation in the frequencies of the other strains as facilitated by within-host mutation.

For both cases, Fig.\ \ref{fig:w-v} illustrates the dependence of the products $w_{0i}v_{0i}$ on the parameter $\alpha$. One can observe that the behavior of the respective terms (that enter the expression for the the sensitivity terms $R^\beta_{1,T}$ and $R^\beta_{1,P}$) closely follows that of $v_{0i}$.

\end{exmp}
\section{Numerical simulation for different scenarios}\label{sec:num}

As for the numerical simulation we consider the following values: $\mu=0.025$, $\gamma=3$, $\xi=5$, $u_{\Tr}=0.4$, and $\beta_{\Cr}=0.25$ (see Example \ref{exmp:1} for an explanation). Similarly to the example, we considered the matrix $A$ as in \eqref{eq:A-ex2} with $\alpha_1\approx 0.97$, $\alpha_2\approx 0.83$, and $\alpha_3= 0.5$, and $\alpha_4=0.25$. At the initial time we set $I_{A1}=0.01$, $S=0.99$, all remaining variables equal to 0.

\begin{figure}[tbh]
\includegraphics[width=0.5\textwidth]{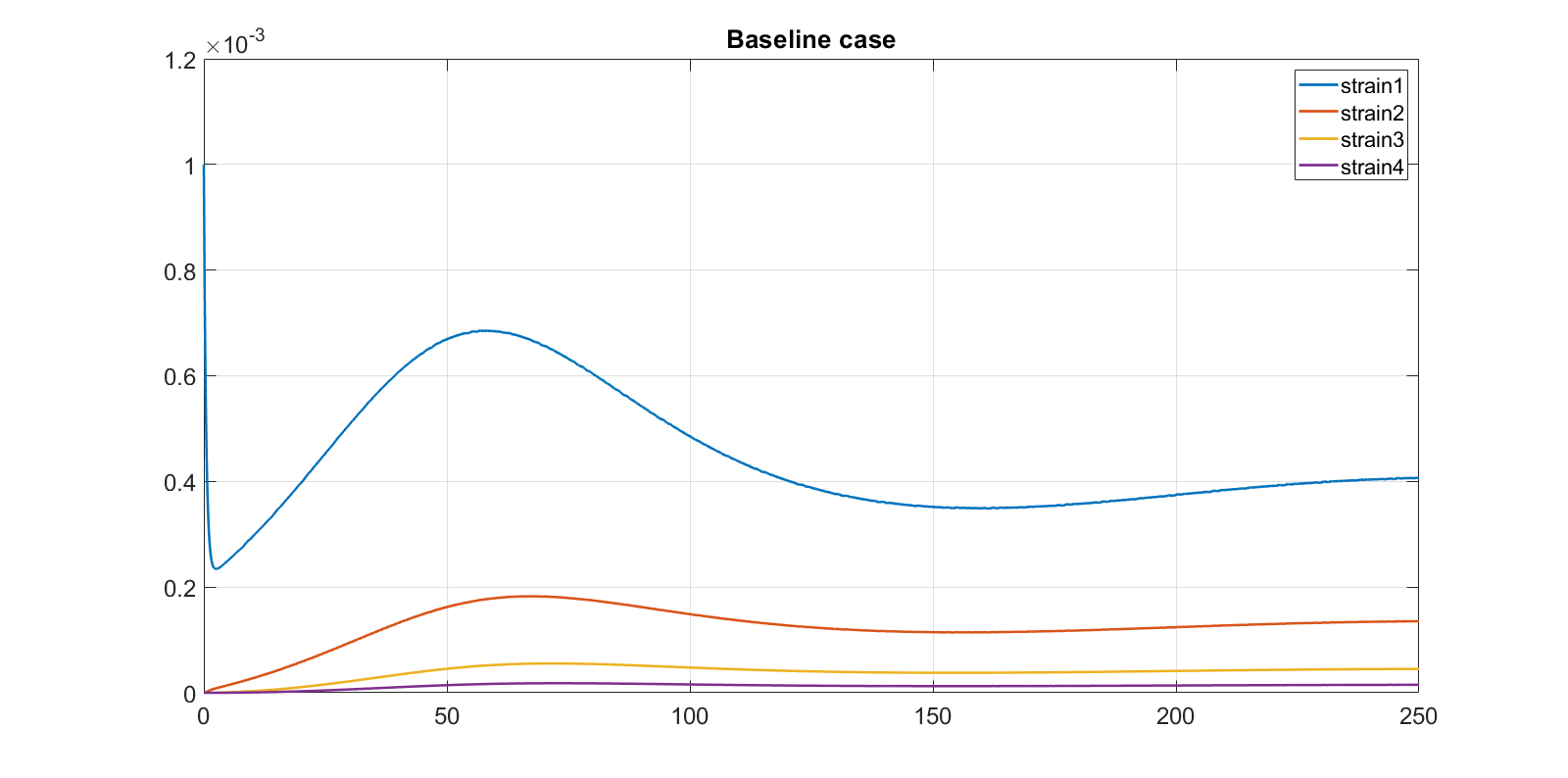}\includegraphics[width=0.5\textwidth]{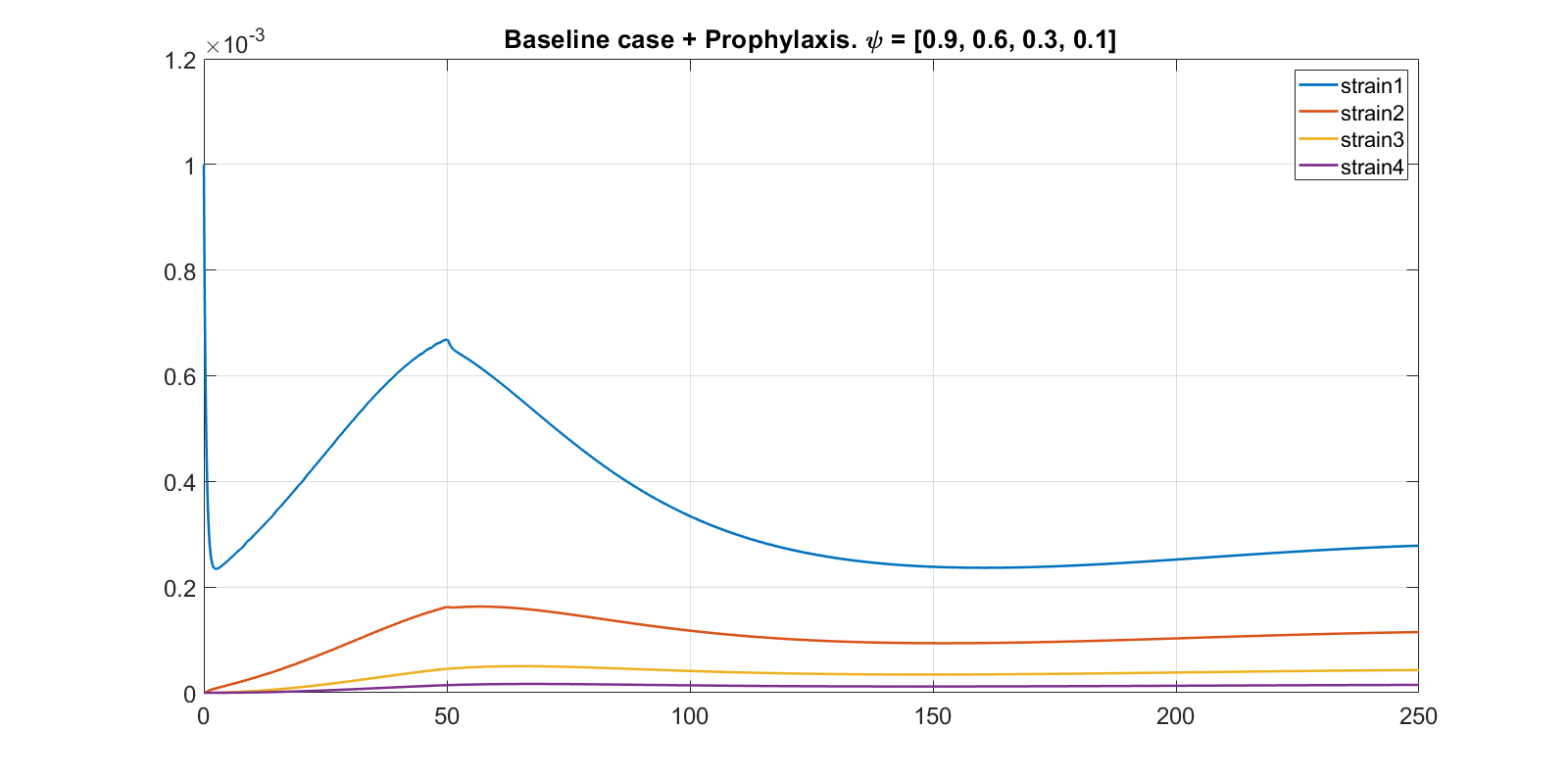}
\hspace*{3cm}a)\hspace*{8cm}b)
\caption{Fractions of acutely infected ($I_{Ai}$) for the baseline model (a) and the baseline model with prophylaxis (b).}\label{fig:f1}
\end{figure}

\paragraph{Propylaxis} First, we simulate the baseline case with treatment and see that the endemic distribution agrees with the assumed values (Fig.\ \ref{fig:f1}a). Next, we turn on prophylaxis at $t=50$ and observe how it changes the process (Fig.\ \ref{fig:f1}b). The respective control $u_P$ was set to $0.1$ and the vector of protections against different virus strains is assumed to be $\psi=[0.9,\,0.6,\,0.3,\,0.1]$.  We see that the endemic frequencies change in the way that the frequency of the first strain decreases (since prophylaxis confers almost total protection against this virus) and the frequency of the remaining strains increases. The (normalized) proportion of different strains in the population of acutely infected is $[0.616,\,    0.255,\, 0.096,\, 0.033]$. For $u_P=0.2$ this effect becomes even more pronounced. The respective frequencies are $[0.547,\, 0.288,\, 0.121,\, 0.043]$.  On the other hand, as $u_P$ grows, the total fraction of infected individuals decreases and asymptotically approaches zero.

Finally, we consider quite a radical setup in which prophylaxis confers total protection against two first virus strains and no protection against the remaining two strains. For $u_P=0.2$, the respective frequencies are $[0.527,\, 0.277,\, 0.144,\, 0.053]$. Comparing with the previous result we see that the difference is rather minute. This implies that while imperfect prophylaxis leads to some increase in the frequencies of the viruses that evade it, this increase is rather restricted. The main reason is that when prophylaxis covers a small fraction of the population it does not create sufficient evolutionary pressure, while when it increases it eventually contributes to the complete eradication of the disease. This result is potentially very encouraging as new prevention methods for HIV based on administration of broadly neutralizing antibodies are predicted to have highly differential levels of protection to diverse viral panels \cite{Wagh:18}. Although further work is needed to explore the potential of selective prophylactic agents to cause strain-level selection in populations in the context of within-host mutation.  

\begin{figure}
\includegraphics[width=0.5\textwidth]{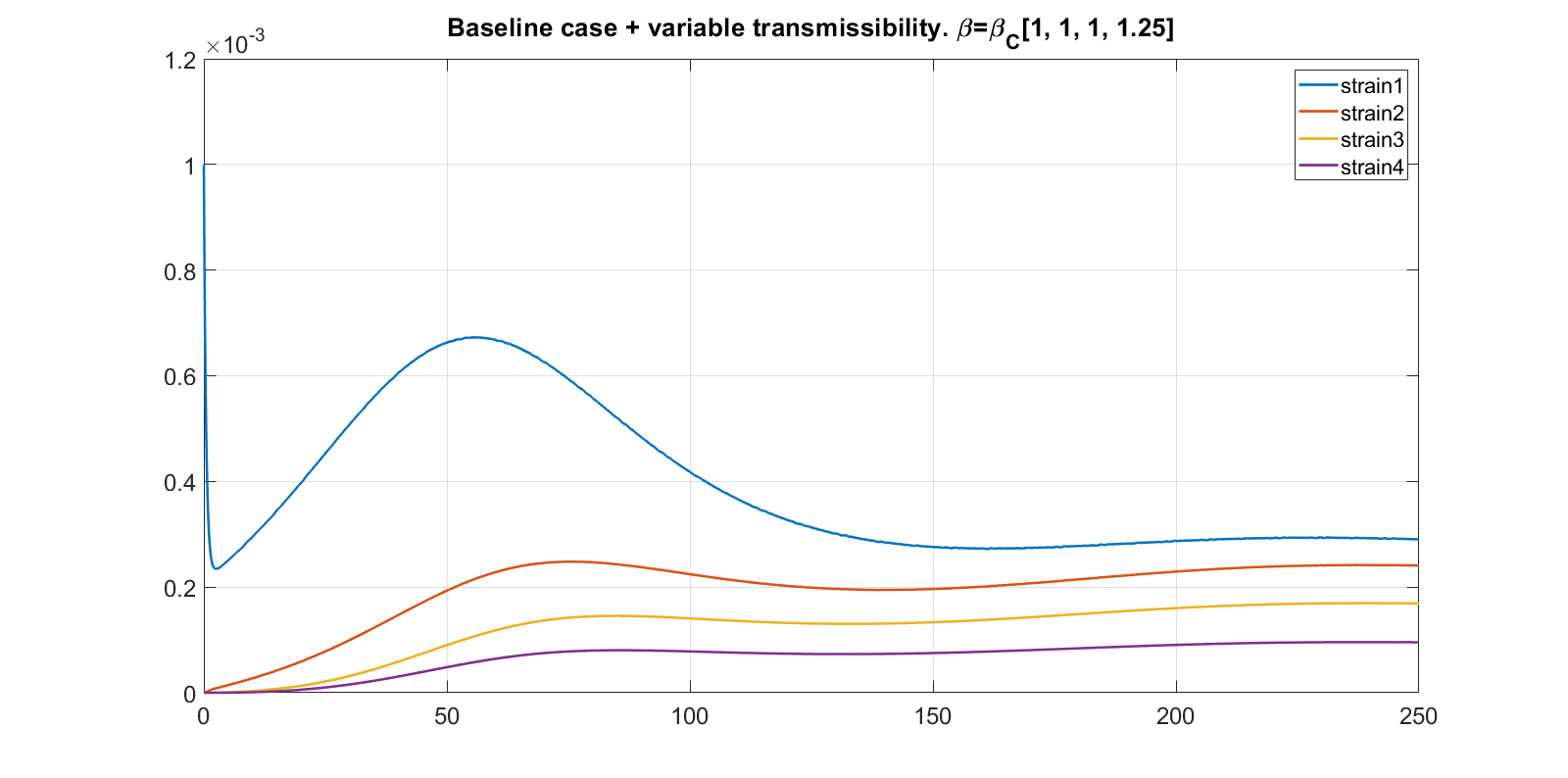}\includegraphics[width=0.5\textwidth]{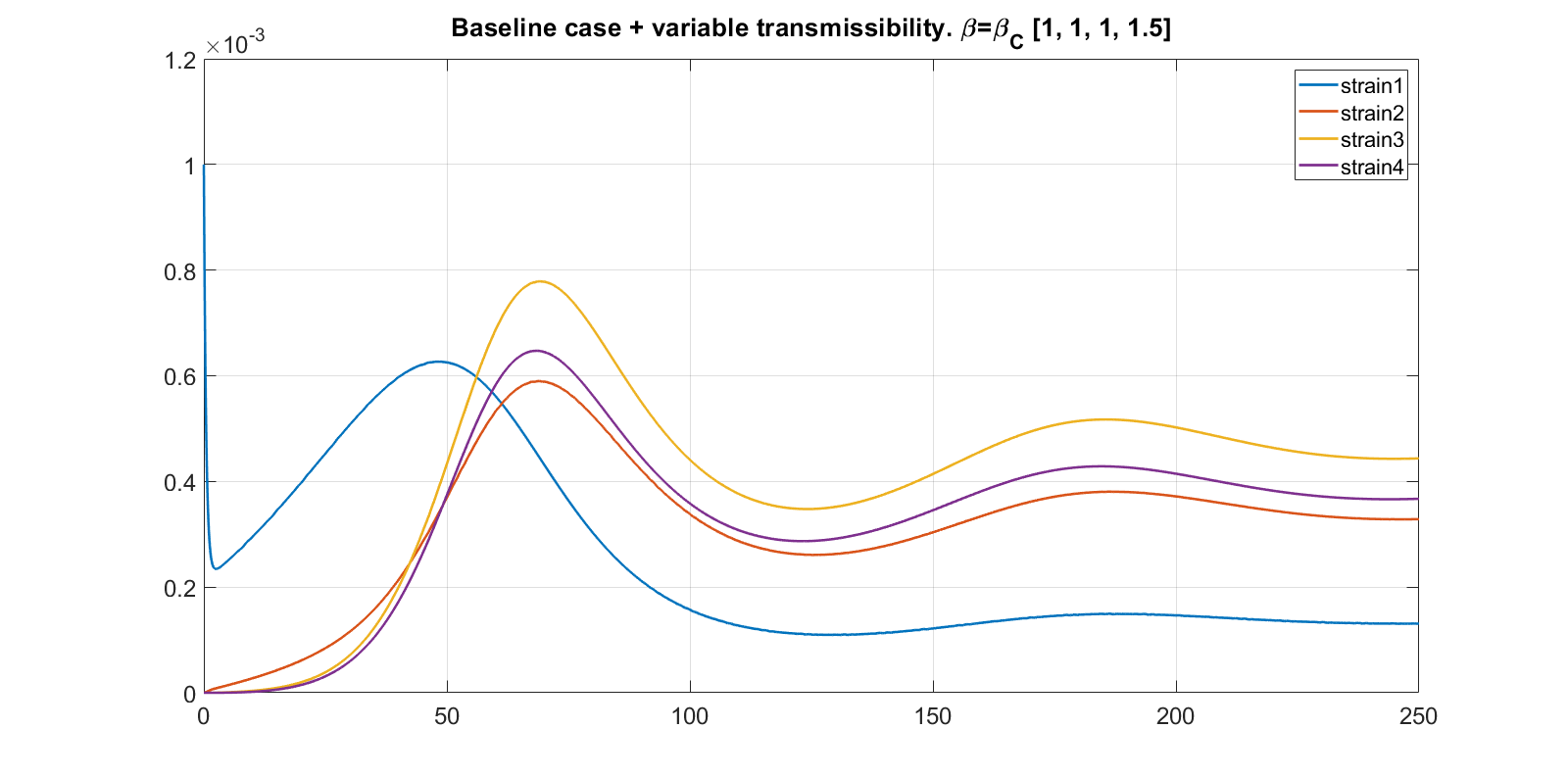}
\hspace*{3cm}a)\hspace*{8cm}b)
\caption{Fractions of acutely infected ($I_{Ai}$) for the model with variable transmissibility.}\label{fig:f2}
\end{figure}

\paragraph{Variable transmissibility} The next experiment is aimed at estimating the effect of variable transmissibility of the final distribution of the virus strains. We assumed that the transmissibility of the fourth strain is either $1.25$ or $1.5$ times larger than the transmissibility of other strains. The results turn out to be quite impressive (see Fig.\ \ref{fig:f2}). While for the first case (Fig.\ \ref{fig:f2}a) the increase in transmissibility has led to a substantial shift in the endemic frequencies: $[0.361,\, 0.304,\, 0.214,\, 0.121]$, the second case is characterized by a complete reshuffle of the strains (Fig.\ \ref{fig:f2}b). It should also be noted that an increase in the transmissibility of one strain not only led to an increase of its endemic frequency, but also resulted in an increase in the endemic frequencies of other strains! While the total fraction of acutely infected at the endemic equilibrium is equal to $\sum_i I_{\Ar i}=7.72\cdot 10^{-4}$ in the first case, it increases to $13.0\cdot 10^{-4}$ in the second case -- almost a doubling! The respective endemic fractions are $[0.103,\, 0.259,\, 0.35,\, 0.289]$.

\begin{figure}
\centerline{\includegraphics[width=0.75\textwidth]{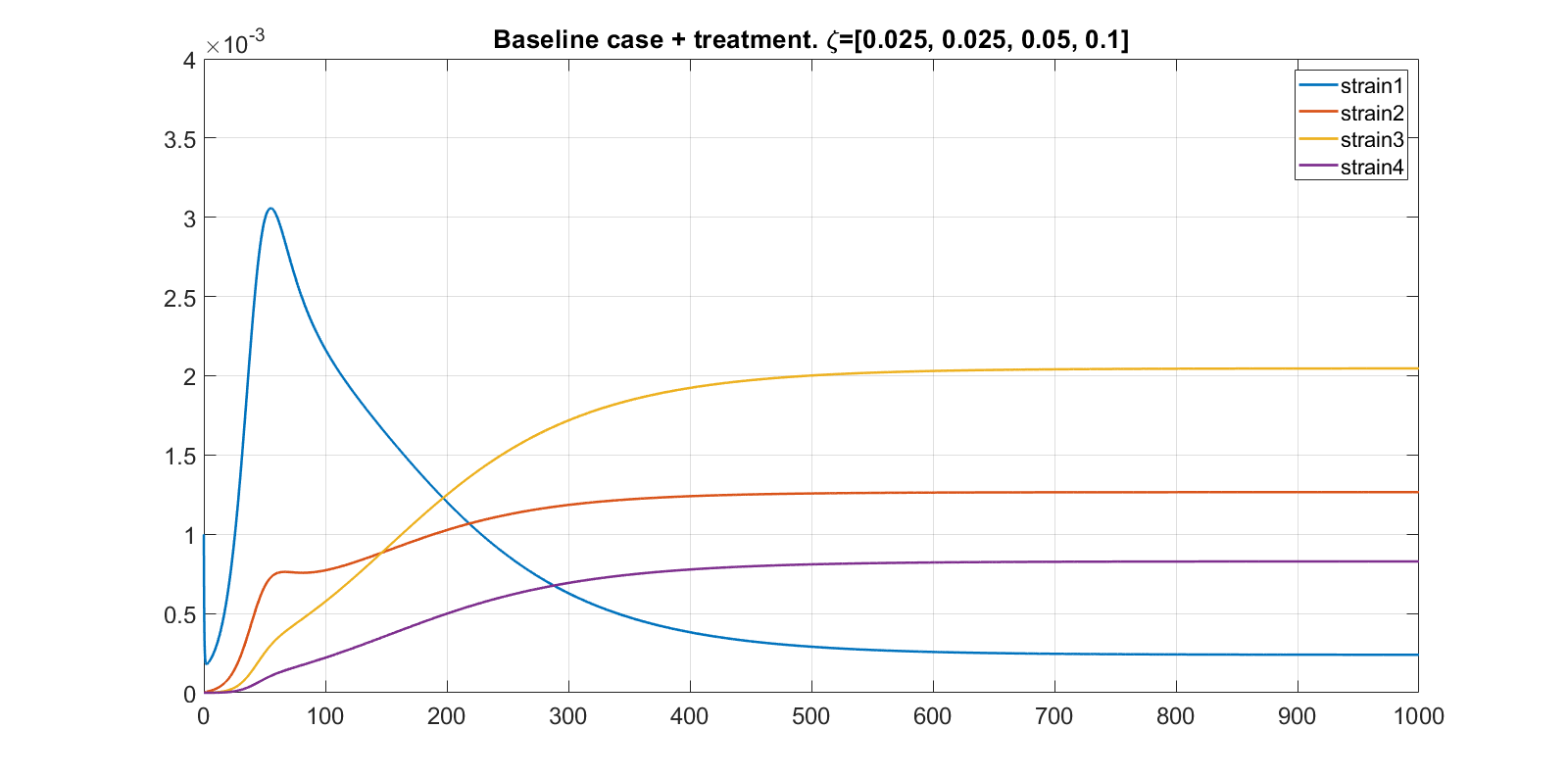}}
\caption{Fractions of acutely infected ($I_{Ai}$) for the model with imperfect treatment.}\label{fig:f3}
\end{figure}

\paragraph{Imperfect treatment} Now we consider the case when there is a chance of treatment failure. We set the respective rates to be rather small: $\zeta=[0.025,\,0.025,\,0.05,\,0.1]$ while keeping all remaining parameters as in the baseline case. We also assume that there is no treatment. The result turns out to be quite surprising: not only the endemic frequencies reshuffle, but also the total proportion of acutely infected increases dramatically, see Fig.\ \ref{fig:f3}. The endemic frequencies are $[0.055,\, 0.289,\, 0.467,\, 0.189]$ and the total fraction of acutely infected is $44.0\cdot 10^{-4}$.

\section*{Conclusions}
In this paper, we described two models of joint evolutionary and epidemiological dynamics of a viral pathogen. While the first, baseline model did not take into account the phenotypic variability of the virus, the extended model addressed the within-host evolution among multiple phenotypes characterized by variable contagiousness, resistance to prophylactic measures, and resistance to therapeutic measures. We presented an analytic expression for the controlled basic reproduction number for both cases and carried out sensitivity analysis of the derived expression to the changes of the control actions. It turned out that the sensitivity coefficients $R_1^{\Tr}$ and $R_1^{\Prm}$ have a straightforward interpretation that can be used when assessing the relative efficacy of the controls. Further, we characterized the endemic equilibria for the baseline model and an extension thereof and shown that a sole assumption of variable transmissibility of different virus strains can lead to wide variations in the endemic distribution of the respective strains. Finally, we carried out a numerical analysis aimed at analyzing the effects of phenotypic diversity of virus strains on the population level dynamics and distribution of different virus strains wihtin the population.

\section*{Appendix A. Necessary ingredients from matrix algebra}

In this appendix, we present some facts about non-negative and stochastic matrices that will be used in the sequel. The interested reader can find a thorough treatment of non-negative matrices, in particular the Perron-Frobenius theory in \cite[Ch.\ 8]{Meyer:00}. The theory of stochastic matrices within the context of Markov chains is detailed in \cite{Kemeny:76}.

\paragraph{Non-negative matrices} A matrix $M$ is said to be {\em non-negative} ({\em positive}), denoted by $M\succeq 0$ ($M\succ 0$), if it is element-wise non-negative (positive). The matrix $M$ is said to be {\em reducible} if there exists a permutation matrix $P$ such that the conjugated matrix $PMP^\top$ has a block upper-triangular form. Otherwise the matrix $M$ is said to be {\em irreducible}. The matrix $M$ is {\em primitive} if it is non-negative and there exists $k\in \mathbb{N}$ such that $M^k\succ 0$. A non-negative irreducible matrix is primitive if at least one diagonal element is non-zero. 

For irreducible non-negative matrices, there exists a real eigenvalue, called {\em dominant} that is equal to the spectral radius of the matrix. The corresponding left and right dominant eigenvectors are positive. This result follows from the celebrated Perron-Frobenius theorem \cite[Ch.\ 8]{Meyer:00}.
In a reducible case, the above results should be substantially weakened to remain true. In particular, there can be multiple eigenvalues corresponding to the spectral radius $r$ and the respective eigenvectors are merely non-negative, rather than positive. If a non-negative matrix $M$ is reducible, it can be transformed to the {\em normal form}, which corresponds to a block upper diagonal matrix such that the diagonal blocks are irreducible.

\paragraph{Stochastic matrices} A non-negative matrix $Q$ is said to be {\em column stochastic} ({\em row stochastic}) is all its columns (rows) sum to 1. The notions of (ir)reducibility, primitivity and the Perron-Frobenius theoren can be extended to stochastic matrices in a straightforward manner. Below we mention several properties that are specific for stochastic matrices.  

A stochastic matrix is typically used to describe the transition structure of a Markov chain. The spectral radius of a stochastic matrix is equal to 1. The respective normalized right eigenvector $v$ is called the {\em stationary distribution} of the respective Markov chain, i.e., $Qv=v$. Here, normalization means that the components of $v$ must sum to 1.  If the stochastic matrix is irreducible, then due to the Perron-Frobenius theorem the stationary distribution is unique and component-wise positive. Finally, we present the result on computing the stationary distribution of an irreducible stochastic matrix. This is a version of the Markov chain tree theorem \cite{Gursoy:13}, formulated using the results from matrix theory (cf. \cite{Wicks_PhD}).
\begin{thm}\label{thm:L} Given an $[n\times n]$ irreducible column stochastic matrix $Q$ , the $i$th element of the right dominant eigenvector of $Q$ is defined as the $i$th principal minor of the corresponding Laplacian $\Lambda=Q-E_n$: $$w_i=[\Lambda]_{i,i}.$$
\end{thm}
\begin{pf}
We have $Qw=w$, which is equivalent to $(Q-E_n)w=\Lambda w=0$. That is, $w$ is the eigenvector corresponding to the zero eigenvalue of $\Lambda$ or, alternatively, $w\in \ker(\Lambda)$. By the Perron-Frobenius theorem, the eigenspace associated with the dominant eigenvalue of $Q$, and hence, the kernel of $\Lambda$ is one-dimensional. 

By the definition of the adjugate, we have $\adj(\Lambda)\Lambda=\Lambda\adj(\Lambda)=\det(\Lambda)E_n=\mathbf{0}_n$. This means, in particular, that the columns of $\adj(\Lambda)$ are linearly dependent and proportional to the stationary distribution $v$.

By transposing the first expression we get $\Lambda^\top\adj(\Lambda)^\top=0$. The columns of $\Lambda$ and hence, the rows of $\Lambda^\top$ sum to 0, which implies that the kernel of $\Lambda^\top$ is a one-dimensional space spanned by $\mathbf{1}_{[n\times 1]}$. This means that each column of $\adj(\Lambda)^\top$ has the form $\col_i\left(\adj(\Lambda)^\top\right)=(-1)^{i+i}[\Lambda]_{i,i}\cdot \mathbf{1}_{[n\times 1]}=[\Lambda]_{i,i}\cdot \mathbf{1}_{[n\times 1]}$. Respectively, each column of $\adj(\Lambda)$ has the form  $\col_i\left(\adj(\Lambda)\right)=\begin{pmatrix}[\Lambda]_{1,1},&\dots,[\Lambda]_{n,n}\end{pmatrix}$. This concludes the proof.\qed
\end{pf}


\section*{Appendix B. Proofs}

\begin{pf}[Theorem \ref{thm_R01}] The Jacobian matrix of \eqref{eq:sys1} evaluated at the DFE has the form 
\begin{equation}\label{eq:J1}\frac{DF}{DX}\bigg|_{X=X_{DFE}}=\begin{bmatrix}
(\xi\beta_{\Cr} - (\gamma + \mu))\Er_n& \beta_{\Cr} \Ar& \mathbf{0}_{[n\times 1]}& \mathbf{0}_{[n\times 1]}\\[5pt]
\gamma\, \Er_n& -( u+\mu)\,\Er_n&\mathbf{0}_{[n\times 1]}&\mathbf{0}_{[n\times 1]}\\[5pt]
\mathbf{0}_{[1\times n]}& u \mathbf{1}_{[1\times n]}&-\mu &0\\[5pt]
-\xi\beta_{\Cr} \mathbf{1}_{[1\times n]}&-\beta_{\Cr} \mathbf{1}_{[1\times n]}&0&-\mu
\end{bmatrix}\end{equation}
We observe that the stability of \eqref{eq:J1} is determined by the eigenvalues of its $[2n\times 2n]$ leading submatrix. As a side remark, we mention that this implies that in our case the computation of $R_0$ requires considering both $I_A$ and $I_{\Cr}$ as infected states (cf. the discussion at the end of Sec. 2 in \cite{JTB:19}). Following the standard procedure, we split the respective submatrix in two thus obtaining
$$\begin{bmatrix}
\xi\beta_{\Cr}\, \Er_n& \beta_{\Cr} A\\[5pt]
\mathbf{0}_{[n\times n]}& \mathbf{0}_{[n\times n]}
\end{bmatrix}+
\begin{bmatrix}
- (\gamma + \mu)\Er_n& \mathbf{0}_{[n\times n]}\\[5pt]
\gamma\, \Er_n& -( u+\mu)\,\Er_n
\end{bmatrix}=F-V.$$
The basic reproductive number is defined as the spectral radius of the product $FV^{-1}$, i.e., $R_0=\rho(FV^{-1})$. Using the standard result on the block matrix inversion we get
\begin{equation}\label{eq:invV}V^{-1} = \frac{1}{(\gamma+\mu)(u_{\Tr}+\mu)}\begin{bmatrix} (u_{\Tr}+\mu) \Er_n &0 \\ \gamma\Er_n & (\gamma+\mu)\Er_n \end{bmatrix},\end{equation}
The product $FV^{-1}$ is equal to
$$\frac{\beta_{\Cr}}{(\gamma+\mu)(u_{\Tr}+\mu)}\begin{bmatrix} \xi(u_{\Tr}+\mu) \Er+ \gamma \Ar&\gamma \Ar \\ \mathbf{0}_n & \mathbf{0}_n \end{bmatrix}$$
and hence, $R_0$ is found as the spectral radius of the $[n\times n]$ matrix $(\gamma+\mu)^{-1}(u_{\Tr}+\mu)^{-1}\beta_{\Cr}\left[\xi(u_{\Tr}+\mu) \Er+ \gamma \Ar\right]$. We use the well known facts that if $\lambda$ is an eigenvalue of the matrix $M$, then $a\lambda$ is an eigenvalue of the matrix $aM$ and $k+\lambda$ is an eigenvalue of the matrix $[k\Er+M]$ for any $\alpha\in\R$, $k\in\R$. This implies that $R_0=\frac{\beta_{\Cr}\gamma}{(\gamma+\mu)(u_{\Tr}+\mu)}\left(\frac{\xi(u_{\Tr}+\mu)}{\gamma}+\rho(A)\right)$. Finally, since $A$ is a column stochastic matrix, it holds that $\rho(A)=1$ and hence, we obtain \eqref{eq:R0}.\qed\end{pf}
\begin{pf}[Theorem \ref{thm_R02}] The Jacobian matrix of \eqref{eq:sys2} evaluated at the DFE has the form 
\begin{multline}\label{eq:J3}\frac{DF}{DX}\bigg|_{X=X_{DFE}}=\\
\begin{bmatrix}
\xi \bar\Psi \Br_{\Cr} - (\gamma + \mu)\Er_n& \bar\Psi \Br_{\Cr} \Ar& \mathbf{0}_n& \mathbf{0}_{[n\times 1]}& \mathbf{0}_{[n\times 1]}\\[5pt]
\gamma\, \Er_n& -( u_{\Tr}+\mu)\,\Er_n&\Zr&\mathbf{0}_{[n\times 1]}& \mathbf{0}_{[n\times 1]}\\[5pt]
\mathbf{0}_n& u_{\Tr} \Er_n&-\mu \Er_n-\Zr&\mathbf{0}_{[n\times 1]}&\mathbf{0}_{[n\times 1]}\\[5pt]
-\xi\mathbf{1}_{[1\times n]}\Br_{\Cr}S_{DFE}&-\mathbf{1}_{[1\times n]}\Br_{\Cr} \Ar S_{DFE}&\mathbf{0}_{[1\times n]}&-(\mu+u_{\Prm})&\delta\\
-\xi\mathbf{1}_{[1\times n]}(\Er_n-\Psi)\Br_{\Cr}S_{DFE}&-\mathbf{1}_{[1\times n]}(\Er_n-\Psi)\Br_{\Cr} \Ar S_{DFE}&\mathbf{0}_{[1\times n]}&u_{\Prm}&-(\delta+\mu)
\end{bmatrix},\end{multline}
where $\bar\Psi=\Er_n-P_{DFE}\Psi$.

The Jacobian \eqref{eq:J3} is a block lower-triangular matrix, whose bottom-right $[2\times 2]$ block is a negated $\mathrm{M}$-matrix and hence Hurwitz. Thus stability of the DFE is determined by the eigenvalues of the leading $[3n\times 3n]$ submatrix. As a side remark, we mention that this implies that the $T_i$ compartments must be considered as infected. That differs from what we observed in the baseline case and emphasizes the importance of the right choice of the infected compartments (see the discussion at the end of Section 2 in \cite{JTB:19}).

 Following the standard procedure, we split the respective submatrix in two:
$$\begin{bmatrix}
\xi \bar\Psi \Br_{\Cr}& \bar\Psi \Br_{\Cr} \Ar& \mathbf{0}_n\\[5pt]
\mathbf{0}_n& \mathbf{0}_n&\mathbf{0}_n\\[5pt]
\mathbf{0}_n& \mathbf{0}_n&\mathbf{0}_n
\end{bmatrix}+\begin{bmatrix}
-(\gamma + \mu)\Er_n& \mathbf{0}_n& \mathbf{0}_n\\[5pt]
\gamma\, \Er_n& -( u_{\Tr}+\mu)\,\Er_n&\Zr\\[5pt]
\mathbf{0}_n& u_{\Tr} \Er_n&-\mu \Er_n-\Zr
\end{bmatrix}=F-V$$
A complete expression for the inverse of $V$ is rather bulky. However, we note that since $R_0$ is computed as the spectral radius of $FV^{-1}$, we need to compute only those blocks of the inverse that enter the leading $[n\times n]$ submatrix of the product $FV^{-1}$. So, we write
$$V^{-1}=
\frac{1}{\gamma + \mu}\begin{bmatrix}
 \Er_n& \mathbf{0}_n& \mathbf{0}_n\\[5pt]
\frac{\gamma}{\mu}\, \Delta(u_{\Tr})& *&*\\[5pt]
*& *&* 
\end{bmatrix},$$
where we used asterisks to denote the blocks that are not relevant to our problem. The diagonal matrix $\Delta(\mathbf{u}_{\Tr})$ is defined as
$$\Delta(u_{\Tr})=(\Zr +(\mu+u_{\Tr}) \Er_n)^{-1}(\Zr+\mu\Er_n).$$
Finally we compute $R_0$ to be
$$R_0(u_{\Tr},u_{\Prm})=\frac{1}{(\gamma+\mu)\mu}\,\rho\left(\xi\mu \bar\Psi \Br_{\Cr}+\gamma\bar\Psi \Br_{\Cr} \Ar \Delta(\mathbf{u}_{\Tr})\right)
=\frac{\bar\beta_{\Cr}(\gamma+\xi\mu)}{(\gamma+\mu)\mu}\,\rho\left(\bar\Br_{\Cr}\bar\Psi\frac1{\gamma+\xi\mu}\left(\xi\mu\Er_n+\gamma \Ar\Delta(\mathbf{u}_{\Tr})\right)\right),
$$
which is the equation \eqref{eq:R02}. Note that $\bar\Br_{\Cr}$ and $\bar\Psi$ are diagonal matrices and therefore commute.\qed\end{pf}
\begin{pf}[Theorem \ref{thm:R0-beta}] The proof is similar to the proof of Theorem 3.4\footnote{Note that there is a typo in Eq. (10) in \cite{JTB:19}. The correct expression is written next to Eq. (12) in {\em op.cit.}} in \cite{JTB:19} and hence will only be sketched. First, we note that $R^\beta_0=R^\beta_0(0,0)=\frac{\bar\beta_{\Cr}(\gamma+\xi\mu)}{(\gamma+\mu)\mu}\rho\left(\bar \Br_{\Cr}\bar \Ar\right)$. Computation of $R^\beta_{1,\Tr}$ and $R^\beta_{1,\Prm}$ coefficients boils down to computing partial derivatives of $R^\beta_0(u_{\Tr},u_{\Prm})$ w.r.t. either $u_{\Tr}$ or $u_{\Prm}$ at $u_{\Tr}=u_{\Prm}=0$. Using the same approach as in \cite{JTB:19}, we get
\begin{equation*}\begin{aligned}
R^\beta_{1,\Tr}={ }&\frac{\bar\beta_{\Cr}(\gamma+\xi\mu)}{(\gamma+\mu)\mu} w_0^\top Q(0)R'(0)v_0,\\
R^\beta_{1,\Prm}={ }&\frac{\bar\beta_{\Cr}(\gamma+\xi\mu)}{(\gamma+\mu)\mu} w_0^\top Q'(0)R(0)v_0.
\end{aligned}\end{equation*}
Noting that $Q(0)=\bar\Br_{\Cr}$, $R(0)=\bar\Ar$, $Q'(0)=\frac{d}{du_{\Prm}}Q(u_{\Prm})\bigg|_{u_{\Prm}=0}\!=-\frac{1}{\delta+\mu}\bar \Br_{\Cr}\Psi$ and $R'(0)=\frac{d}{du_{\Tr}}R(u_{\Tr})\bigg|_{u_{\Tr}=0}\!=-\frac{\gamma}{\gamma+\xi\mu}\Ar(\Zr+\mu\Er_n)^{-1}$ we obtain
\begin{subequations}\begin{align}
R^\beta_{1,\Tr}={ }&-\frac{\bar\beta_{\Cr}\gamma}{(\gamma+\mu)\mu} w_0^\top \bar\Br_{\Cr}\Ar(\Zr+\mu\Er_n)^{-1}v_0,\label{eq:R1T}\\
R^\beta_{1,\Prm}={ }&-\frac{\bar\beta_{\Cr}(\gamma+\xi\mu)}{(\delta+\mu)(\gamma+\mu)\mu} w_0^\top \bar \Br_{\Cr}\Psi\bar\Ar v_0.\label{eq:R1P}
\end{align}\end{subequations}
The expressions \eqref{eq:R1T} and \eqref{eq:R1P} can be further transformed using the fact that $w_0$ and $v_0$ are the left and the right eigenvectors of $\bar\Br_{\Cr}\bar\Ar$ corresponding to the spectral radius of this matrix and expressing $\rho\left(\bar\Br_{\Cr}\bar\Ar\right)$ through the basic reproduction number $R_0^\beta$ as shown below.
\begin{multline*}R^\beta_{1,\Tr}=-\frac{\bar\beta_{\Cr}}{(\gamma+\mu)\mu} w_0^\top \bar\Br_{\Cr}[(\gamma+\xi\mu)\bar\Ar - \xi\mu \Er_n](\Zr+\mu\Er_n)^{-1}v_0\\
=-\frac{\bar\beta_{\Cr}(\gamma+\xi\mu)}{(\gamma+\mu)\mu} w_0^\top \rho(\bar\Br_{\Cr}\bar\Ar)(\Zr+\mu\Er_n)^{-1}v_0 + \frac{\bar\beta_{\Cr}\xi}{(\gamma+\mu)} w_0^\top \bar\Br_{\Cr}(\Zr+\mu\Er_n)^{-1}v_0\\
=-R_0^\beta w_0^\top (\Zr+\mu\Er_n)^{-1}v_0 + \frac{\bar\beta_{\Cr}\xi}{(\gamma+\mu)} w_0^\top \bar\Br_{\Cr}(\Zr+\mu\Er_n)^{-1}v_0
=-w_0^\top\left[R_0^\beta \Er_n - \frac{\xi}{(\gamma+\mu)} \Br_{\Cr}\right](\Zr+\mu\Er_n)^{-1}v_0
\end{multline*}
$$R^\beta_{1,\Prm}=-\frac{\bar\beta_{\Cr}(\gamma+\xi\mu)}{(\delta+\mu)(\gamma+\mu)\mu} w_0^\top \Psi\bar\Br_{\Cr}\bar\Ar v_0=-\frac{\bar\beta_{\Cr}(\gamma+\xi\mu)}{(\delta+\mu)(\gamma+\mu)\mu} w_0^\top \Psi\rho\left(\bar\Br_{\Cr}\bar\Ar\right) v_0=-R_0^\beta \frac{1}{(\delta+\mu)} w_0^\top \Psi v_0$$\qed
\end{pf}

\begin{pf}[Theorem \ref{thm2}] When computing the endemic equilibrium we let the equilibrium value of $S$ be equal to some (not yet known) value $S^*$. The respective equilibrium values of $I_{Ai}$ and $I_{Ci}$ are found as the solution of the following system of $2n$ algebraic equations:
\begin{equation}\label{eq:I-eq2}\begin{aligned}
0={ }&\xi\beta_{\Cr} S^* \Ir_A + \beta_{\Cr} S^* A \Ir_{\Cr} -(\gamma +\mu) \Ir_A\\
0={ }&\gamma \Ir_{A}-( u + \mu) \Ir_{\Cr},\\
\end{aligned}\end{equation}
From the first equation we get
$$\Ir^*_A=  \frac{\beta_{\Cr} S^*}{(\gamma+\mu) - \xi\beta_{\Cr} S^*} \Ar \Ir^*_{\Cr}.
$$
Expressing $\Ir^*_{\Cr}$ from the second equation of \eqref{eq:I-eq2} and substituting into the above equation we obtain the expression for $\Ir_A^*$:
\begin{equation}\label{eq:I*}\left[\frac{((\gamma+\mu) - \xi\beta_{\Cr} S^*)(u_{\Tr}+\mu)}{\beta_{\Cr}\gamma S^*}\Er_n -  \Ar\right] \Ir^*_A=\Gamma \Ir^*_A = 0,
\end{equation}
that is $\Ir_A^*$ belongs to the kernel of the matrix $\Gamma=\frac{((\gamma+\mu) - \xi\beta_{\Cr} S^*)(u_{\Tr}+\mu)}{\beta_{\Cr}\gamma S^*}\Er_n -  \Ar$. The matrix $\Ar$ is non-singular, thus $\Gamma$ has a non-trivial kernel only if the factor in front of $\Er_n$ is equal to one of the eigenvalues of $\Ar$. The solution $\Ir^*_A$ would be then equal (up to a positive factor) to the corresponding eigenvector. However, as follows from the Perron-Frobenius theorem, the only positive eigenvector corresponds to its dominant eigenvalue, which is equal to 1 for a stochastic matrix. Any other eigenvector would contain negative components which contradict the assumption that all system's states are non-negative. This implies that the equilibrium value of $S$ must satisfy the equation $\frac{((\gamma+\mu) - \xi\beta_{\Cr} S^*)(u_{\Tr}+\mu)}{\beta_{\Cr}\gamma S^*}-1=0$, whence we obtain
\begin{equation}\label{eq:S*}S^*=\frac{(\gamma+\mu)(u_{\Tr}+\mu)}{\xi\beta_{\Cr} (u_{\Tr}+\mu) + \beta_{\Cr}\gamma}=\frac{1}{R_0}.\end{equation}
The equilibrium solution $\Ir^*_A$ corresponds to the right dominant eigenvector of the column stochastic matrix $\Ar$. Since this eigenvector is defined up to the multiplication by a positive scalar, we can further specify it using the following argument. Let $i_{A\Sigma}$ be the sum of all components of $\Ir_A$, i.e., $i_{A\Sigma}=\sum_{j=1}^n I^*_{Ai}$. Similarly, we write $i_{C\Sigma}=\frac{\gamma}{(u_{\Tr}+\mu)}i_{A\Sigma}$ and subsequently, we get $T^*=\frac{\gamma u}{(u_{\Tr}+\mu)\mu}i_{A\Sigma}$. Since all states sum to 1, we have the following equation
$$i_{A\Sigma}+\frac{\gamma}{(u_{\Tr}+\mu)}i_{A\Sigma}+\frac{\gamma u}{(u_{\Tr}+\mu)\mu}i_{A\Sigma}+\frac{1}{R_0}=1,$$
whence we get
\begin{equation}\label{eq:sums}i_{A\Sigma}=\frac{\mu}{(\gamma+\mu)}\left(1-\frac{1}{R_0}\right),\; i_{C\Sigma}=\frac{\gamma \mu}{(\gamma+\mu)(u_{\Tr}+\mu)}\left(1-\frac{1}{R_0}\right),\; T^*= \frac{\gamma  u}{(\gamma+\mu)(u_{\Tr}+\mu)}\left(1-\frac{1}{R_0}\right).\end{equation}
Let $v$ be the (normalized) dominant eigenvector of $A$ such that $\sum_{i=1}^n v_i=1$. Multiplying the components of $v$ with the respective factors we obtain the expressions for $I_{Ai}$ and $I_{Ci}$. 

Finally, we observe that $R_0<1$ implies that the respective components of the endemic equilibrium state turn negative which implies that there is no admissible endemic equilibrium. This concludes the proof.\qed\end{pf}
\section*{References}
\bibliographystyle{plainnat}
\bibliography{multistrain}

\begin{thebibliography}{21}
\providecommand{\natexlab}[1]{#1}
\providecommand{\url}[1]{\texttt{#1}}
\expandafter\ifx\csname urlstyle\endcsname\relax
  \providecommand{\doi}[1]{doi: #1}\else
  \providecommand{\doi}{doi: \begingroup \urlstyle{rm}\Url}\fi

\bibitem[Bianco et~al.(2009)Bianco, Shaw, and Schwartz]{Bianco:09}
Simone Bianco, Leah~B. Shaw, and Ira~B. Schwartz.
\newblock Epidemics with multistrain interactions: {T}he interplay between
  cross immunity and antibody-dependent enhancement.
\newblock \emph{Chaos: An Interdisciplinary Journal of Nonlinear Science},
  19\penalty0 (4):\penalty0 043123, 2009.
\newblock \doi{10.1063/1.3270261}.

\bibitem[Breban et~al.(2010)Breban, Drake, and Rohani]{Breban:10}
Romulus Breban, John~M. Drake, and Pejman Rohani.
\newblock A general multi-strain model with environmental transmission:
  {I}nvasion conditions for the disease-free and endemic states.
\newblock \emph{Journal of Theoretical Biology}, 264\penalty0 (3):\penalty0
  729--736, 2010.
\newblock \doi{10.1016/j.jtbi.2010.03.005}.

\bibitem[De~Leenheer and Pilyugin(2008)]{Leenheer:08}
Patrick De~Leenheer and Sergei~S. Pilyugin.
\newblock {Multistrain virus dynamics with mutations: a global analysis}.
\newblock \emph{Mathematical Medicine and Biology: A Journal of the IMA},
  25\penalty0 (4):\penalty0 285--322, 2008.
\newblock \doi{10.1093/imammb/dqn023}.

\bibitem[Gromov et~al.(2017)Gromov, Bulla, Serea, and Romero-Severson]{MMB:17}
Dmitry Gromov, Ingo Bulla, Oana~Silvia Serea, and Ethan~O. Romero-Severson.
\newblock Numerical optimal control for {HIV} prevention with dynamic budget
  allocation.
\newblock \emph{Mathematical Medicine and Biology: A Journal of the IMA},
  35\penalty0 (4):\penalty0 469--491, 2017.
\newblock \doi{10.1093/imammb/dqx015}.

\bibitem[Gromov et~al.(2019)Gromov, Bulla, and Romero-Severson]{JTB:19}
Dmitry Gromov, Ingo Bulla, and Ethan~O. Romero-Severson.
\newblock Systematic evaluation of the population-level effects of alternative
  treatment strategies on the basic reproduction number.
\newblock \emph{Journal of Theoretical Biology}, 462:\penalty0 381--390, 2019.
\newblock \doi{10.1016/j.jtbi.2018.11.029}.

\bibitem[Gursoy et~al.(2013)Gursoy, Kirkland, Mason, and Sergeev]{Gursoy:13}
Buket~Benek Gursoy, Steve Kirkland, Oliver Mason, and Sergei Sergeev.
\newblock On the {M}arkov chain tree theorem in the {M}ax algebra.
\newblock \emph{Electronic Journal of Linear Algebra}, 26\penalty0
  (1):\penalty0 2, 2013.

\bibitem[Kemeny and Snell(1976)]{Kemeny:76}
John~G. Kemeny and J.~Laurie Snell.
\newblock \emph{Finite {M}arkov chains}.
\newblock Undergraduate Texts in Mathematics. Springer, second edition, 1976.

\bibitem[Kooi et~al.(2013)Kooi, Aguiar, and Stollenwerk]{Kooi:13}
Bob~W. Kooi, Ma\'ira Aguiar, and Nico Stollenwerk.
\newblock Bifurcation analysis of a family of multi-strain epidemiology models.
\newblock \emph{Journal of Computational and Applied Mathematics},
  252:\penalty0 148--158, 2013.
\newblock \doi{10.1016/j.cam.2012.08.008}.

\bibitem[Kryazhimskiy et~al.(2007)Kryazhimskiy, Dieckmann, Levin, and
  Dushoff]{Kryazhimskiy:07}
Sergey Kryazhimskiy, Ulf Dieckmann, Simon~A. Levin, and Jonathan Dushoff.
\newblock On state-space reduction in multi-strain pathogen models, with an
  application to antigenic drift in influenza {A}.
\newblock \emph{PLoS Computational Biology}, 3\penalty0 (8):\penalty0 e159,
  2007.
\newblock \doi{10.1371/journal.pcbi.0030159}.

\bibitem[Kucharski et~al.(2016)Kucharski, Andreasen, and Gog]{Kucharski:16}
Adam~J. Kucharski, Viggo Andreasen, and Julia~R. Gog.
\newblock Capturing the dynamics of pathogens with many strains.
\newblock \emph{Journal of mathematical biology}, 72\penalty0 (1-2):\penalty0
  1--24, 2016.
\newblock \doi{10.1007/s00285-015-0873-4}.

\bibitem[Lythgoe et~al.(2013)Lythgoe, Pellis, and Fraser]{Lythgoe:13}
Katrina~A. Lythgoe, Lorenzo Pellis, and Christophe Fraser.
\newblock Is {HIV} short-sighted? {I}nsights from a multistrain nested model.
\newblock \emph{Evolution}, 67\penalty0 (10):\penalty0 2769--2782, 2013.
\newblock \doi{10.1111/evo.12166}.

\bibitem[Melchjorsen et~al.(2009)Melchjorsen, Matikainen, and Paludan]{MMS:09}
Jesper Melchjorsen, Sampsa Matikainen, and S{\o}ren~R. Paludan.
\newblock Activation and evasion of innate antiviral immunity by herpes simplex
  virus.
\newblock \emph{Viruses}, 1\penalty0 (3):\penalty0 737--759, 2009.
\newblock \doi{10.3390/v1030737}.

\bibitem[Meyer(2000)]{Meyer:00}
Carl~D. Meyer.
\newblock \emph{Matrix analysis and applied linear algebra}.
\newblock SIAM, 2000.

\bibitem[Minayev and Ferguson(2008)]{Minayev:08}
Pavlo Minayev and Neil Ferguson.
\newblock Improving the realism of deterministic multi-strain models:
  implications for modelling influenza {A}.
\newblock \emph{Journal of the Royal Society, Interface}, 6:\penalty0 509--518,
  2008.
\newblock \doi{10.1098/rsif.2008.0333}.

\bibitem[Organization(2019)]{WHO:19}
World~Health Organization.
\newblock {HIV} drug resistance report 2019.
\newblock Technical Report WHO/CDS/HIV/19.21, WHO, 2019.
\newblock URL
  \url{https://www.who.int/hiv/pub/drugresistance/hivdr-report-2019/en/}.

\bibitem[Perelson et~al.(1996)Perelson, Neumann, Markowitz, Leonard, and
  Ho]{Perelson:96}
Alan~S. Perelson, Avidan~U. Neumann, Martin Markowitz, John~M. Leonard, and
  David~D. Ho.
\newblock {HIV}-1 dynamics in vivo: Virion clearance rate, infected cell
  life-span, and viral generation time.
\newblock \emph{Science}, 271\penalty0 (5255):\penalty0 1582--1586, 1996.
\newblock \doi{10.1126/science.271.5255.1582}.

\bibitem[Van~den Driessche and Watmough(2002)]{Driessche:02}
Pauline Van~den Driessche and James Watmough.
\newblock Reproduction numbers and sub-threshold endemic equilibria for
  compartmental models of disease transmission.
\newblock \emph{Mathematical biosciences}, 180\penalty0 (1):\penalty0 29--48,
  2002.

\bibitem[van~der Ventel(2011)]{Ventel:11}
Brandon~I.S. van~der Ventel.
\newblock Analysis of the multistrain asymmetric si model for arbitrary strain
  diversity.
\newblock \emph{Mathematical and Computer Modelling}, 53\penalty0 (5):\penalty0
  1007--1025, 2011.
\newblock \doi{10.1016/j.mcm.2010.11.058}.

\bibitem[Wagh et~al.(2018)Wagh, Seaman, Zingg, Fitzsimons, Barouch, Burton,
  Connors, Ho, Mascola, Nussenzweig, Ravetch, Gautam, Martin, Montefiori, and
  Korber]{Wagh:18}
Kshitij Wagh, Michael~S. Seaman, Marshall Zingg, Tomas Fitzsimons, Dan~H.
  Barouch, Dennis~R. Burton, Mark Connors, David~D. Ho, John~R. Mascola,
  Michel~C. Nussenzweig, Jeffrey Ravetch, Rajeev Gautam, Malcolm~A. Martin,
  David~C. Montefiori, and Bette Korber.
\newblock Potential of conventional \& bispecific broadly neutralizing
  antibodies for prevention of {HIV}-1 subtype {A}, {C} \& {D} infections.
\newblock \emph{PLOS Pathogens}, 14\penalty0 (3):\penalty0 1--24, 03 2018.
\newblock \doi{10.1371/journal.ppat.1006860}.

\bibitem[Wicks(2009)]{Wicks_PhD}
John~R Wicks.
\newblock \emph{An algorithm to compute the stochastically stable
  distributionof a perturbed Markov matrix}.
\newblock PhD thesis, Brown University, 2009.
\newblock URL
  \url{https://cs.brown.edu/research/pubs/theses/phd/2009/wicks.pdf}.

\bibitem[Wikramaratna et~al.(2015)Wikramaratna, Kucharski, Gupta, Andreasen,
  McLean, and Gog]{Wikra:15}
Paul~S. Wikramaratna, Adam Kucharski, Sunetra Gupta, Viggo Andreasen, Angela~R.
  McLean, and Julia~R. Gog.
\newblock Five challenges in modelling interacting strain dynamics.
\newblock \emph{Epidemics}, 10:\penalty0 31--34, 2015.
\newblock \doi{10.1016/j.epidem.2014.07.005}.

\end{thebibliography}

\end{document}